\documentclass[12pt,a4paper,english,superscriptaddress,aps,nofootinbib]{revtex4}
\usepackage{amsmath,amssymb,graphicx}
\makeatletter
\usepackage{babel}
\usepackage[active]{srcltx}
\usepackage{graphicx,color}
\usepackage{changebar}
\usepackage[utf8x]{inputenc}
\usepackage{hyperref}
\hypersetup{
	colorlinks=true,
	urlcolor= blue,
	citecolor=blue,
	linkcolor= blue,
	bookmarks=true,
	bookmarksopen=false,}

\usepackage[T1]{fontenc}
\usepackage{ae}
\usepackage{amsmath}
\usepackage{braket}
\usepackage{mathtools}
\usepackage{slashed}
\usepackage{empheq}
\usepackage{tikz}
\usepackage{multirow}
\usetikzlibrary{decorations.pathmorphing}
\usepackage[caption = false]{subfig}

\usepackage{graphicx}
\usepackage{amsfonts}
\usepackage{amsmath}
\newcommand{\be}{\begin{equation}}
	\newcommand{\ee}{\end{equation}}
\newcommand{\bea}{\begin{eqnarray}}
	\newcommand{\eea}{\end{eqnarray}}

\begin{document}
	\title{Non-minimal coupling of fermion to the torsion in the modied
teleparallel braneworld}
	
\author{A. R. P. Moreira}
\affiliation{Universidade Federal do Cear\'{a} (UFC), Departamento do F\'{i}sica,\\ Campus do Pici, Fortaleza, CE, C. P. 6030, 60455-760, Brazil.}

\author{J. E. G. Silva}
\affiliation{Universidade Federal do Cariri(UFCA), Av. Tenente Raimundo Rocha, \\ Cidade Universit\'{a}ria, Juazeiro do Norte, Cear\'{a}, CEP 63048-080, Brazil.}

\author{C. A. S. Almeida}
\email[]{Email address: carlos@fisica.ufc.br (C. A. S. Almeida).}
\affiliation{Universidade Federal do Cear\'{a} (UFC), Departamento do F\'{i}sica,\\ Campus do Pici, Fortaleza, CE, C. P. 6030, 60455-760, Brazil.}

\begin{abstract}
\vspace{0.5cm}
We study a spin $1/2$ fermion in a teleparallel $f(T)$ domain-wall thick braneworld. By assuming a non-minimal coupling of fermion to the torsion, a geometric alternative to the Yukawa coupling is found. The torsion parameters control the width of the massless Kaluza-Klein mode and the properties of the analogous quantum-potential near the origin. The non-normalized massive fermionic modes are also analyzed.
\end{abstract}

\maketitle
\thispagestyle{empty}

\section{Introduction}

The study of extra-dimensional theories still attracts a lot of attention. One of the first extra dimensions theory was proposed by T. Kaluza \cite{Kaluza1921} and O. Klein \cite{Klein1926}, with the aim of 
unify Einstein's general relativity (GR) and Maxwell's electromagnetism. In the final years of the 20th century, two interesting extra-dimensional theories appeared, characterized by the use of non-compact extra-dimensions. The first one was proposed by N. Arkani-Hamed, S. Dimopoulos, and G.  Dvali \cite{Arkani-Hamed:1998jmv}. The second one was introduced by L. Randall and R. Sundrum \cite{rs,rs2}. The brane model proposed by Randall-Sundrum (RS) aroused a lot of interest and curiosity, as it was able to raise a new viewpoint of spacetime, which made possible new proposals to address a large number of unsolved problems in modern physics. For example, the hierarchy problem and cosmological problem \cite{rs2,cosmologicalconstant}. Also, this new view provides explanations for the nature of dark matter and dark energy \cite{darkmatter}, and enables black hole production at future colliders as a window on quantum gravity \cite{Giddings2001}.

The RS model motivated the emergence of various braneworld scenarios, such as the Gregory-Rubakov-Sibiryakov (GRS) model \cite{Gregory:2000jc}, the universal extra dimension model \cite{Appelquist:2000nn}, the Dvali-Gabadadze-Porrati (DGP) model \cite{Dvali:2000hr}, and the thick brane model \cite{DeWolfe:1999cp,Gremm1999, Csakil,Gremm2000, Dzhunushaliev:2009,Dzhunushaliev:2010, Herrera-Aguilar:2009, Dzhunushaliev:2007, Gogberashvili:2003a,Gogberashvili:2003b}, which has been originated from the domain wall model proposed by V. Rubakov and M. Shaposhnikov \cite{Rubakov:1983bb}. In the thick brane model, the brane could be constructed by the scalar field \cite{Goldberger1999,DeWolfe:1999cp,Gremm1999, Bazeia2008,Dzhunushaliev:2009}, as well as vector fields and spinor fields \cite{Dzhunushaliev:2010,Geng:2015kvs,Dzhunushaliev:2011mm}, and there are also braneworld models without matter fields \cite{Herrera-Aguilar:2009}.

Assuming a warped geometry, the propagation of the fields can be modified, becoming dependent on the bulk curvature. We can cite modification in gravitational field \cite{Csakil}, gauge field \cite{Kehagias}, and fermionic field \cite{Almeida2009}). Also, the localization mechanism employed for matter fields in braneworld scenarios has been the subject of many studies, in particular, the study of fermion localization. The most used method in literature for locating fermions is formulated in a very speculative way, due to the great freedom of choice for the Yukawa coupling term \cite{Almeida2009, RandjbarDaemi2000, Liu2009, Liu2009a, Liu2008, Liu2009b, Liu2008b, Obukhov2002,Ulhoa2016,Dantas2013,Sousa2014,Dantas}. In the context of modified gravity, the localization of the fermion field was studied in a generalization of GR, known as $f(R)$ gravity \cite{Mitra2017, Buyukdag2018, Wang2019}. 

Another modified gravity theory that has attracted a lot of attention recently is the $f(T)$ gravity \cite{ft, Ferraroinflation}, which is an extension of the teleparallel equivalent of general relativity (TEGR).  Similar to TEGR, $f(T)$ gravity assumes that the gravitational interaction is encoded in the torsion, unlike GR where the gravitational interaction is encoded in the curvature \cite{Andrade2001, Aldrovandi,Linder, Bahamonde:2021gfp}. One of the most significant advantages of $f(T)$ gravity has over other modified gravity theories is that it leads to second-order equations of the motion (eom) \cite{ftpalatini}. 

In braneworld scenarios the $f(T)$ gravity has shown significant results, as the emergence of internal structures, changes on the stress-energy tensor, and modifications in gravitational perturbations \cite{Yang2017, ftenergyconditions,tensorperturbations}, as well as changes in the dynamics of fermions \cite{Yang2012}. As a matter of fact, Yang and collaborators \cite{Yang2012}, considered general fermionic Yukawa coupling between one scalar field and spinorial fields for $f(T)$ gravity. It was found that the simplest Yukawa coupling $\overline{\psi}\phi\psi$ allowed left-handed fermions to possess a zero-mode that is localized on the brane. The right-handed fermions, on the other hand, present no zero-mode.

The results obtained in Ref. \cite{Yang2012} encouraged us to investigate the possibility to study the question of localization of fermions in the gravity $f(T)$ for a non-minimal coupling of the fermion to the torsion. In section (\ref{sec1}) we review the main definitions of the teleparallel $f(T)$ theory and build the respective braneworld. Furthermore, we examined the energy density components of the brane. In the section (\ref{sec2}) the main result of the work is presented, i. e., in the fermionic sector of the model, where we investigate how spacetime torsion influences the localization of fermion fields in the brane using a non-minimal coupling with the torsion $g(T)$. Finally, conclusions and additional comments are presented in section (\ref{finalremarks}).

\section{Dynamical Equations and Solutions}
\label{sec1}

The teleparallel gravity uses the Weitzenb\"{o}ck spacetime. Instead of a pseudo-Riemannian spacetime which disregards torsion, the Weitzenb\"{o}ck spacetime disregarding curvature and the relevant connection is the so-called Weitzenb\"{o}ck connection, which is defined in terms of a dynamical vielbein or tetrad fields, $h^{\overline{M}}\ _M$ (instead of the metric) that works like dynamic variables. The vielbein can be considered as an orthonormal basis in the tangent space connected to each spacetime point. The relation of vielbein with the metric field of spacetime is given via
\begin{eqnarray}
g_{MN}=\eta_{\overline{M}\ \overline{N}}h^{\overline{M}}\ _Mh^{\overline{N}}\ _N,
\end{eqnarray}
where $\eta_{\overline{M}\ \overline{N}}$ assume a mostly plus metric signature, i.e., $diag(-1,1,1,1,1)$. The latin indices  ($M , N , Q ,...=0, 1, 2, 3, 4$ ) are related to the bulk, and barred latin indices ($\overline{M}, \overline{N}, \overline{Q},...=0, 1, 2, 3, 4$ ) are related to tangent space.

The Weitzenböck connection is described as
\begin{eqnarray}
\widetilde{\Gamma}^P\ _{NM}=h_{\overline{M}}\ ^P\partial_M h^{\overline{N}}\ _M,
\end{eqnarray}
which leads us to the condition of absolute parallelism $\nabla_\lambda h^{\overline{M}}\ _M\equiv \partial_Q h^{\overline{M}}\ _M-  \widetilde{\Gamma}^P\ _{QM} h^{\overline{M}}\ _P =0$  \cite{Aldrovandi}.
An important characteristic of this connection is that the corresponding spin connection is canceled. The contortion tensor
\begin{eqnarray}
K^P\ _{NM}=\frac{1}{2}\Big( T_N\ ^P\ _M +T_M\ ^P\ _N - T^P\ _{NM}\Big),
\end{eqnarray}
arises from the difference between the connections of Weitzenböck and Levi–Civita \cite{Aldrovandi}
\begin{eqnarray}
\widetilde{\Gamma}^P\ _{NM}= \Gamma^P\ _{NM} + K^P\ _{NM},
\end{eqnarray}
where $\Gamma^P\ _{NM}$ is the Levi-Civita connection of GR, 
remembering that the curvature in the Weitzenböck connection is canceled. The torsion is described in terms of the Weitzenböck connection as \cite{Aldrovandi}
\begin{eqnarray}
T^{P}\  _{NM}= \widetilde{\Gamma}^P\ _{MN}-\widetilde{\Gamma}^P\ _{NM}. 
\end{eqnarray}
By using the torsion and contortion tensors, we describe the dual torsional tensor \cite{Aldrovandi}
\begin{eqnarray}
S_{P}\ ^{NM}=\frac{1}{2}\Big( K^{NM}\ _{P}-\delta^M_P T^{QN}\ _Q+\delta^N_P T^{QM}\ _Q\Big).
\end{eqnarray}
Furthermore, by defining the torsion tensor and its dual, we can construct the torsion scalar
\begin{eqnarray}
T=T^{P}\  _{NM}S_{P}\ ^{NM}.
\end{eqnarray}

Thus, it is possible to introduce a modified theory of gravity considering the gravitational Lagrangian depending on a general function of torsion \cite{Ferraroinflation}. Therefore, we have the gravitational action to $f(T)$ as 

\begin{equation}\label{55.5}
\mathcal{S}=-\frac{1}{4}\int h f(T)d^5x-\int h\Big[\frac{1}{2}\partial^M\phi\partial_M\phi+V(\phi)\Big] d^5x,
\end{equation}
where $h=\sqrt{-g}$ and $c^4/8\pi G=1$, for simplicity. 

In our work, we consider the braneworld scenario, whose metric is
\begin{equation}\label{45.a}
ds^2=e^{2A(y)}\eta_{\mu\nu}dx^\mu dx^\nu+dy^2,
\end{equation}
where $e^{A(y)}$ is the warped factor, which is related to the brane width. A good choice for vielbein, considering the metric (\ref{45.a}), would be
\begin{eqnarray}\label{000098}
h_{\overline{M}}\ ^M=diag(e^A, e^A, e^A, e^A, 1),
\end{eqnarray} 
for they generate gravitational field equations that do not involve any additional constraints on the function $f(T)$ or the scalars $T$ \cite{Ferraro2011us, Tamanini2012}. The torsion scalar is $T=-12A'^2$, where the prime $( ' )$ denotes differentiation with respect to $y$. Thus, with the vielbein ansatz given by Eq. (\ref{000098}), the gravitational field equations are 
\begin{eqnarray}\label{w.0}
\phi''+4A'\phi&=&\frac{d V}{d\phi},\\
\label{w.1}
6A'^2f_T+\frac{1}{4}f&=& \frac{1}{2}\phi^2-V,\\
\label{w.2}
\frac{1}{4}f+\Bigg(\frac{3}{2}A''+6A'^2\Bigg)f_T-36A'^2A''f_{TT}&=&-\frac{1}{2}\phi^2-V,
\end{eqnarray} 
where $f\equiv f(T)$, $f_T\equiv\partial f(T)/\partial T$ and $f_{TT}\equiv\partial^2 f(T)/\partial T^2$ for simplicity. 

Then, we will propose three cases where $f_1(T)=T+kT^{n_1}$, $f_2(T)=n_2 \sinh\left(\frac{T}{n_2}\right)$ and $f_3(T)=n_3 \tanh\left(\frac{T}{n_3}\right)$, where $k$ and $n_{1,2,3}$ are torsion parameters.

 The first case $f_1(T)$ was the first to be proposed in the study of the behavior of the brane in a modified teleparallel gravity $f(T)$ \cite{Yang2012}. The results found in Ref.\cite{Yang2012}, motivated the further study of this model in a braneworld scenario \cite{Menezes,ftnoncanonicalscalar, ftborninfeld,ftmimetic,tensorperturbations}. 
  
In order to explain acceleration of the Universe \cite{Linder}, dark energy \cite{ftgw}, and cosmology \cite{ftenergyconditions}, several $f(T)$ models were studied. On the other hand, at the same time, the interest in studying these models in a brane-world scenario arose as well as \cite{ftnoncanonicalscalar,mirza, ftmimetic}.  These models are exactly our others $f_2(T)$ and  $f_3(T)$ cases mentioned above.

It is difficult to provide an analytical solution for these cases, although the Eqs.(\ref{w.1}) and (\ref{w.2}) form a second-order derived theory.  To overcome this difficulty, we take the ansatz given by Gremm \cite{Gremm1999}, namely 
\begin{eqnarray}\label{20}
e^{2A(y)}=\cosh^{-2p}(\lambda y),
\end{eqnarray}
where the $p$ and $\lambda$ parameters modify the warp variation and determine the width within the brane core, respectively. 

\subsection{Scalar field solution}

In this subsection, we focus on obtaining solutions for the scalar field in the thick brane, considering the three proposed functions of torsion. 
\subsubsection{$f_1(T)=T+kT^{n_1}$}

For $f_1(T)$, the thick brane solutions are investigated in Ref. \cite{Yang2012}. The equations are 
\begin{eqnarray}\label{q.5}
\phi'^2(y)&=&\frac{3}{2}p\lambda^2\mathrm{sech}^2(\lambda y)-\frac{(-12)^{n_1}}{8p}\Big\{kn_1(2n_1-1)\mathrm{csch}^2(\lambda y)\Big[p\lambda\tanh(\lambda y)\Big]^{2n_1}\Big\},\\
\label{q.55}
V(\phi(y))&=&\frac{3}{2}p\lambda^2\mathrm{sech}^2(\lambda y)-9[p\lambda\tanh(\lambda y)]^2+\frac{(-12)^{n_1}}{8p}\Bigg\{k\Big[4(2n_1+1)p\nonumber\\&-&n_1(2n_1-1)\mathrm{csch}^2(\lambda y)\Big] \Big[p\lambda\tanh(\lambda y)\Big]^{2n_1}\Bigg\}.
\end{eqnarray}

\begin{figure}[ht!]
\begin{center}
\begin{tabular}{ccc}
\includegraphics[height=5cm]{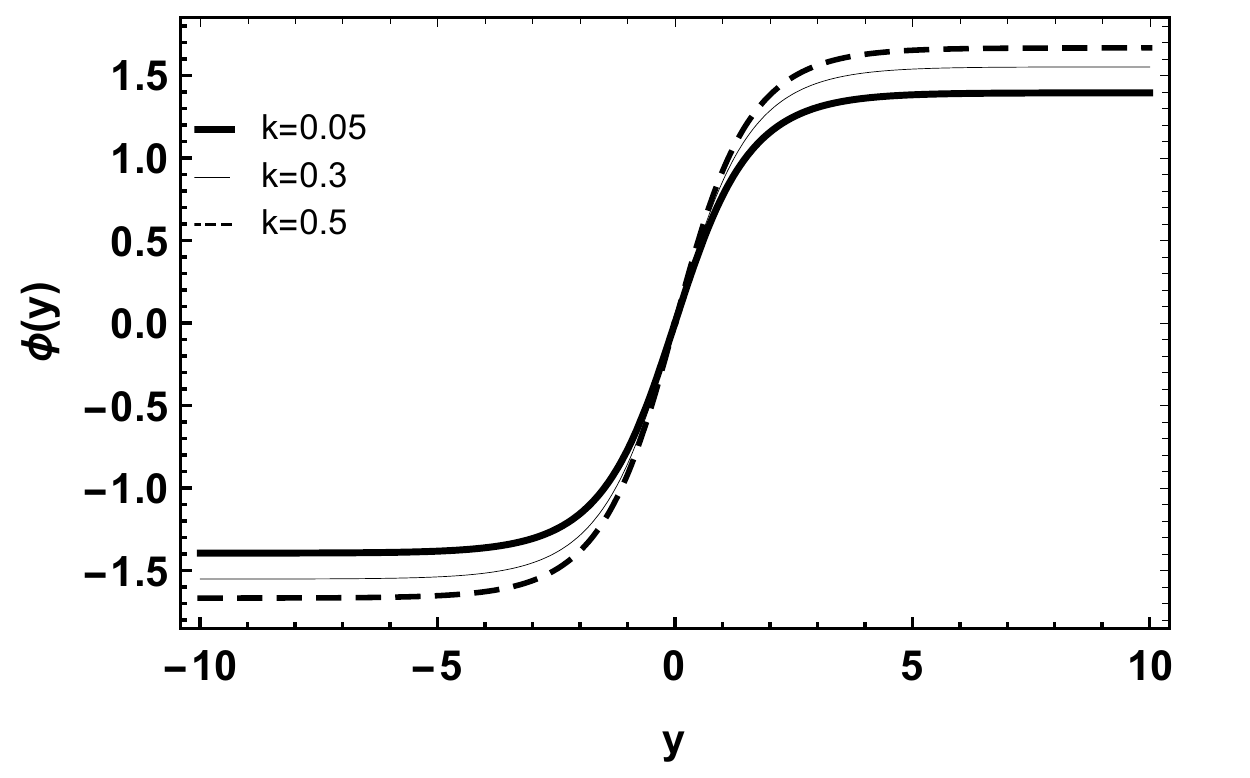} 
\includegraphics[height=5cm]{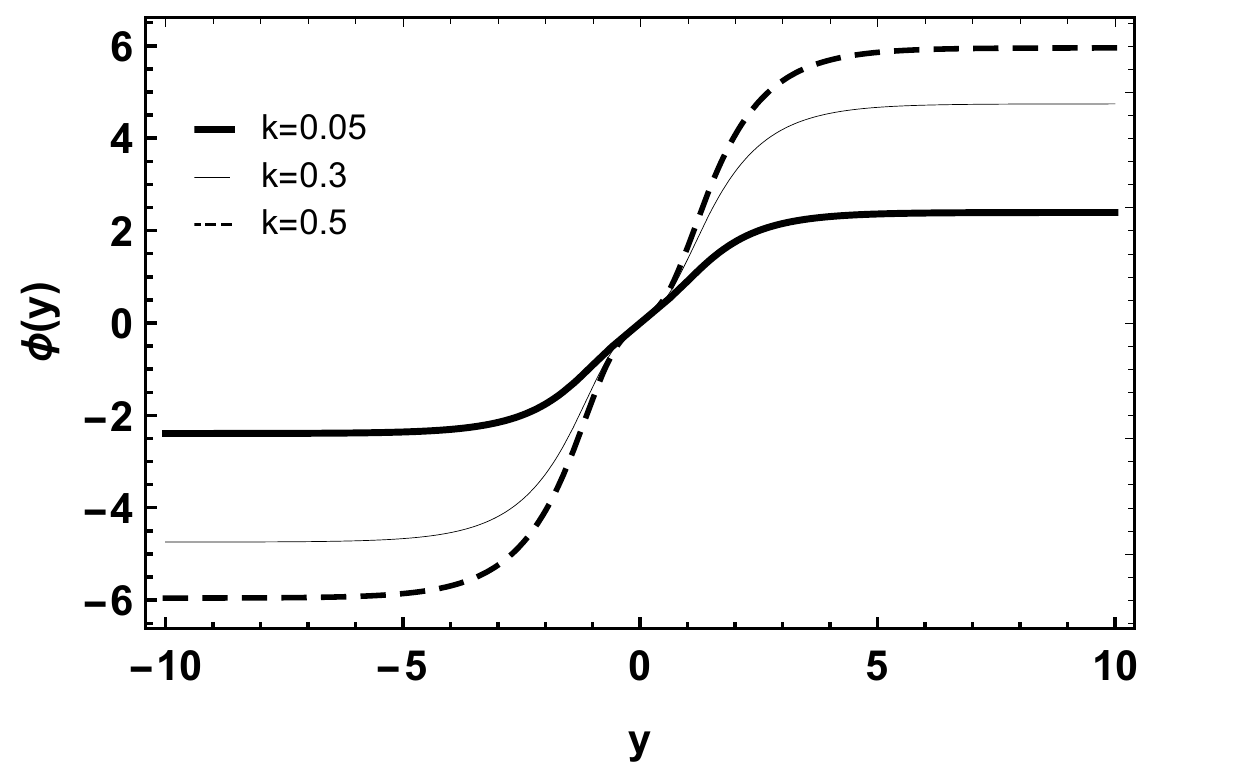}\\
(a) \hspace{8 cm}(b)
\end{tabular}
\end{center}
\caption{The shape of the scalar field $\phi(y)$ for $f_1(T)$, where $p=0.4$ and $\lambda=1$. (a)  $n_1=1$. (b) $n_1=3$. 
\label{fig01}}
\end{figure}

From Eq.(\ref{q.5}),  we obtain the solution of the scalar field $\phi(y)$, which assumes a constant value $\phi_{c_1}$ asymptotically. As we can see, Eq.(\ref{q.5}) is not so simple to solve. We plotted the $\phi(y)$ field for $f_1(T)$ in Fig.\ref{fig01}. For $n_1=1$ configuration  (Fig.\ref{fig01}$a$)  we have a kink solution, whereas for  $n_1=3$ configuration, we have a double-kink solution (Fig.\ref{fig01}$b$). When we increase the value of $n_1$ for both cases, we also increase the value of $\phi_{c_1}$. From equation (\ref{q.55}), we get the asymptotic values of the potential, which takes the form of a cosmological constant
\begin{eqnarray}
\Lambda\equiv V(\phi\rightarrow\pm\phi_{c_1})=3(p\lambda)^2+(-4)^{n_1-1}k(2n_1-1)[3(p\lambda)^2]^{n_1},
\end{eqnarray}
and its derivative with respect to the field $\partial V(\phi\rightarrow\pm\phi_{c_1})/\partial \phi=0$. This ensures that our model makes physical sense. 

\subsubsection{$f_2(T)=n_2 \sinh\left(\frac{T}{n_2}\right)$}

For $f_2(T)$, we have
\begin{eqnarray}\label{q.6}
\phi'^2(y)&=&\frac{3}{2}p\lambda^2\mathrm{sech}^2(\lambda y)\Bigg\{\cosh\Big[\frac{12}{n_2}\Big(p\lambda\tanh(\lambda y)\Big)^2\Big]\nonumber\\
&+&\frac{24}{n}\sinh\Big[\frac{12}{n_2}\Big(p\lambda\tanh(\lambda y)\Big)^2\Big]
\Big(p\lambda\tanh(\lambda y)\Big)^2\Bigg\},\\
\label{q.66}
V(\phi(y))&=&\frac{3}{4}p\lambda^2\mathrm{sech}^2(\lambda y)\Big[1+4p-4p\cosh(2\lambda y)\Big]\cosh\Big[\frac{12}{n_2}\Big(p\lambda\tanh(\lambda y)\Big)^2\Big]\nonumber\\
&-& \frac{1}{4n_2} \Big[n_2^2-72p^3\lambda^4\mathrm{sech}^2(\lambda y)\tanh(\lambda y)^2\Big]\sinh\Big[\frac{12}{n_2}\Big(p\lambda\tanh(\lambda y)\Big)^2\Big].
\end{eqnarray}

\begin{figure}[ht!]
\begin{center}
\begin{tabular}{ccc}
\includegraphics[height=5cm]{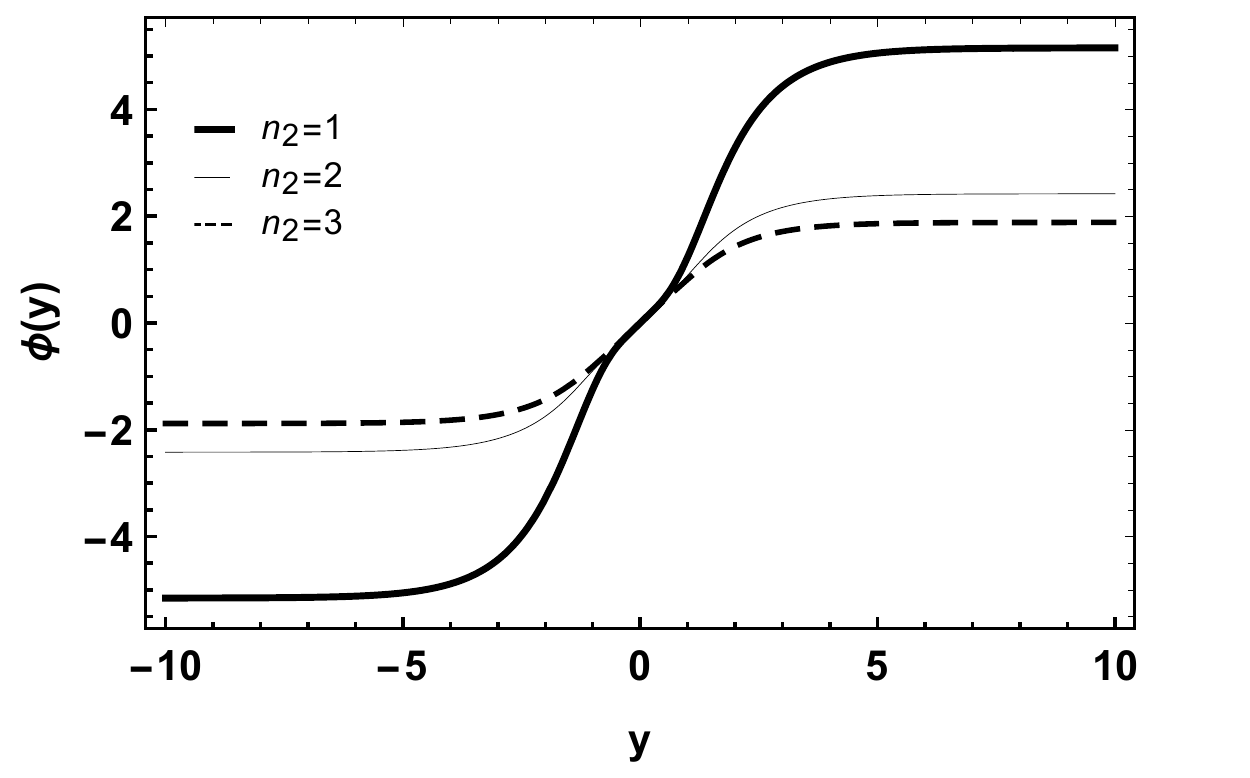} 
\end{tabular}
\end{center}
\caption{The shape of the scalar field $\phi(y)$ for $f_2(T)$, where $p=0.4$ and $\lambda=1$.  
\label{fig02}}
\end{figure}

We obtain the solution of the scalar field $\phi(y)$ numerically solving Eq.(\ref{q.6}). We plotted the $\phi(y)$ field for $f_2(T)$ in Fig.\ref{fig02}, which is clearly a double-kink solution. The scalar $\phi(y)$ assumes a constant value $\phi_{c_2}$ asymptotically, where increasing  the value of $n_2$, we decrease the value of $\phi_{c_2}$. With equation (\ref{q.66}), we get the asymptotic values of the potential, which takes the form of a cosmological constant
\begin{eqnarray}
\Lambda\equiv V(\phi\rightarrow\pm\phi_{c_2})=6(p\lambda)^2\cosh\Big[\frac{12(p\lambda)^2}{n_2}\Big]-\frac{n_2}{4}\sinh\Big[\frac{12(p\lambda)^2}{n_2}\Big],
\end{eqnarray}
and its derivative with respect to the field $\partial V(\phi\rightarrow\pm\phi_{c_2})/\partial \phi=0$, ensuring that our model makes physical sense for this choice of $f(T)$ as well.

\subsubsection{$f_3(T)=n_3 \tanh\left(\frac{T}{n_3}\right)$}

\begin{figure}[ht!]
\begin{center}
\begin{tabular}{ccc}
\includegraphics[height=5cm]{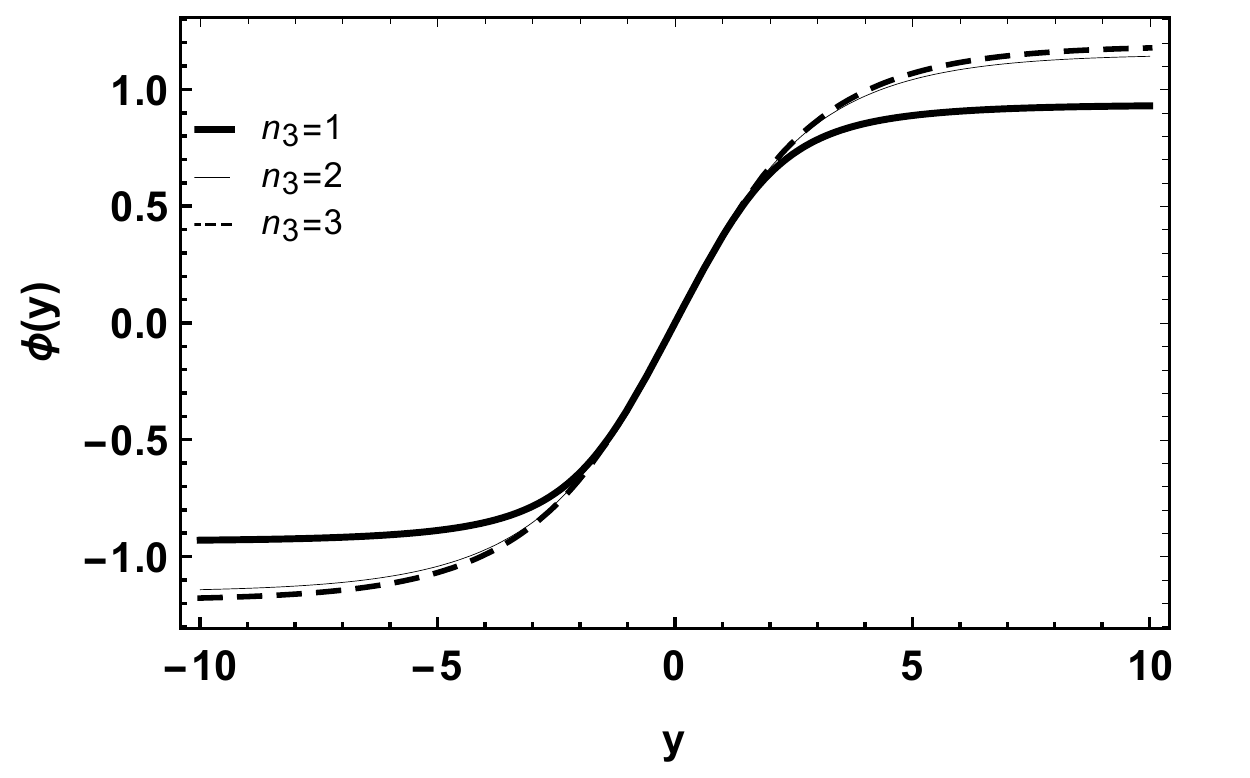} 
\end{tabular}
\end{center}
\caption{ The shape of the scalar field $\phi(y)$ for $f_3(T)$, where $p=0.4$ and $\lambda=0.5$. 
\label{fig03}}
\end{figure}

In turn, for $f_3(T)$, we have
\begin{eqnarray}\label{q.7}
\phi'^2(y)&=&\frac{3}{2}p\lambda^2\mathrm{sech}^2(\lambda y)\Bigg\{\mathrm{sech}^2\Big[\frac{12}{n_3}\Big(p\lambda\tanh(\lambda y)\Big)^2\Big]\nonumber\\
&+&\frac{48}{n_3}\mathrm{sech}^2\Big[\frac{12}{n_3}\Big(p\lambda\tanh(\lambda y)\Big)^2\Big]\tanh\Big[\frac{12}{n_3}\Big(p\lambda\tanh(\lambda y)\Big)^2\Big]
\Big(p\lambda\tanh(\lambda y)\Big)^2\Bigg\},\\
\label{q.77}
V(\phi(y))&=&\frac{3p\lambda^2}{4n_3}\Bigg\{\Big[n_3-48\Big(p\lambda\tanh(\lambda y)\Big)^2\tanh\Big[\frac{12}{n_3}\Big(p\lambda\tanh(\lambda y)\Big)^2\Big]\Big]\mathrm{sech}^2(\lambda y)\nonumber\\&-&8pn_3\tanh(\lambda y)^2\Bigg\}\mathrm{sech}^2\Big[\frac{12}{n_3}\Big(p\lambda\tanh(\lambda y)\Big)^2\Big]\nonumber\\&-&\frac{n_3}{4}\tanh\Big[\frac{12}{n_3}\Big(p\lambda\tanh(\lambda y)\Big)^2\Big].
\end{eqnarray}
We repeat the procedure and obtain the solution of the scalar field $\phi(y)$ numerically solving Eq.(\ref{q.7}). In Fig.\ref{fig03}, we plotted the $\phi(y)$ field for $f_3(T)$, which is clearly a kink solution, that tends to a constant value $\phi_{c_3}$ asymptotically, where increasing the value of $n_3$, the value of $\phi_{c_3}$ increases too. From Eq.(\ref{q.77}), we get the asymptotic values of the potential, which takes the form of a cosmological constant
\begin{eqnarray}
\Lambda\equiv V(\phi\rightarrow\pm\phi_{c_3})=6(p\lambda)^2\mathrm{sech}\Big[\frac{12(p\lambda)^2}{n_3}\Big]-\frac{n_3}{4}\tanh\Big[\frac{12(p\lambda)^2}{n_3}\Big],
\end{eqnarray}
and its derivative with respect to the field $\partial V(\phi\rightarrow\pm\phi_{c_3})/\partial \phi=0$, ensuring again that our model makes physical sense for this choice of $f(T)$.

\subsection{Energy conditions}

In this subsection, we analyze the profile of energy density and the thick brane splitting process. We make a variable change in the form
\begin{eqnarray}
dz=e^{-A(y)}dy,
\end{eqnarray}
where we rewrite the metric (\ref{45.a}) as $ds^2=e^{2A}(\eta^{\mu\nu}dx^\mu dx^\nu+dz^2)$. So, we can find the energy densities in the brane for the three cases $f_{1,2,3}(T)$.

\subsubsection{$f_1(T)=T+kT^{n_1}$}

In this first case, we have
\begin{eqnarray}\label{21}
\rho(z)=\frac{1}{8pz^2\lambda^2}\Big\{\zeta[1-(1+2p)z^2\lambda^2]+k(2{n_1}-1)(-1)^{n_1}\zeta^{n_1}[({n_1}+2p)z^2\lambda^2-{n_1}]\Big\},
\end{eqnarray}
where $\zeta= 12p^2z^2\lambda^4(1+z^2\lambda^2)^{p-2}$. 

\begin{figure}[ht!]
\begin{center}
\begin{tabular}{ccc}
\includegraphics[height=5cm]{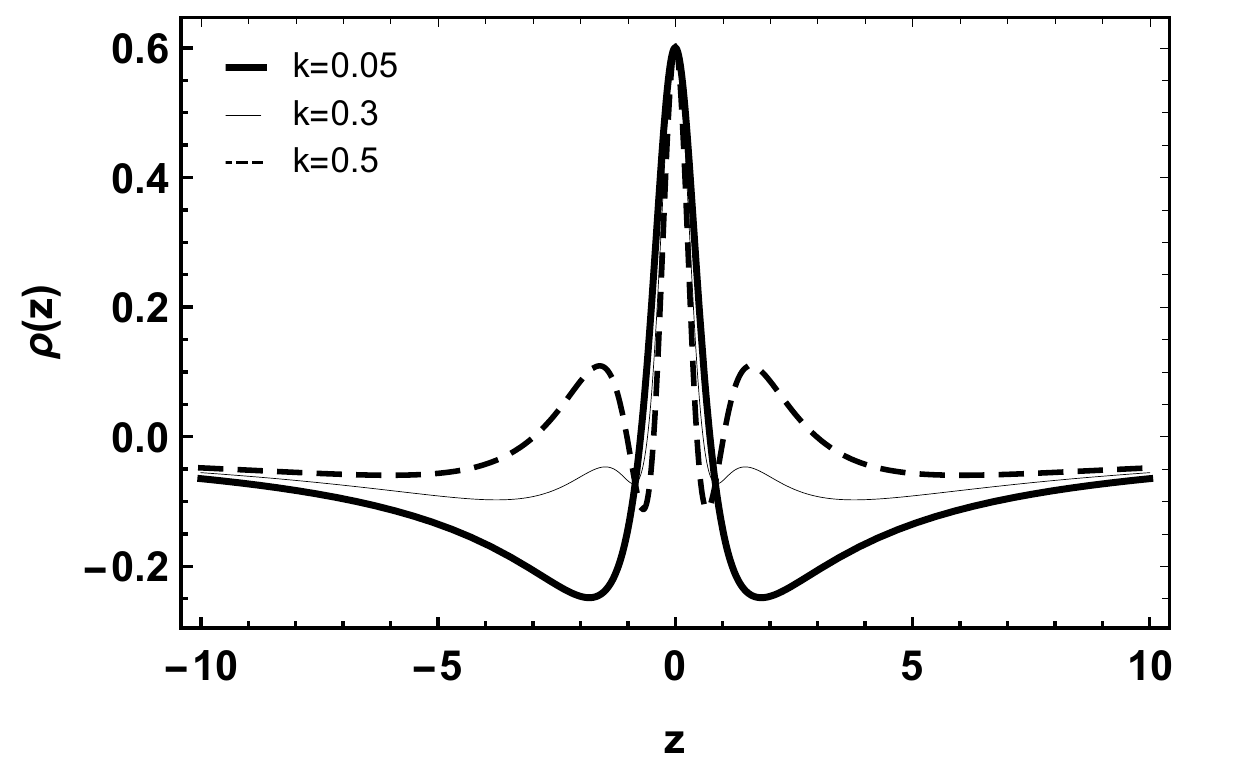} 
\includegraphics[height=5cm]{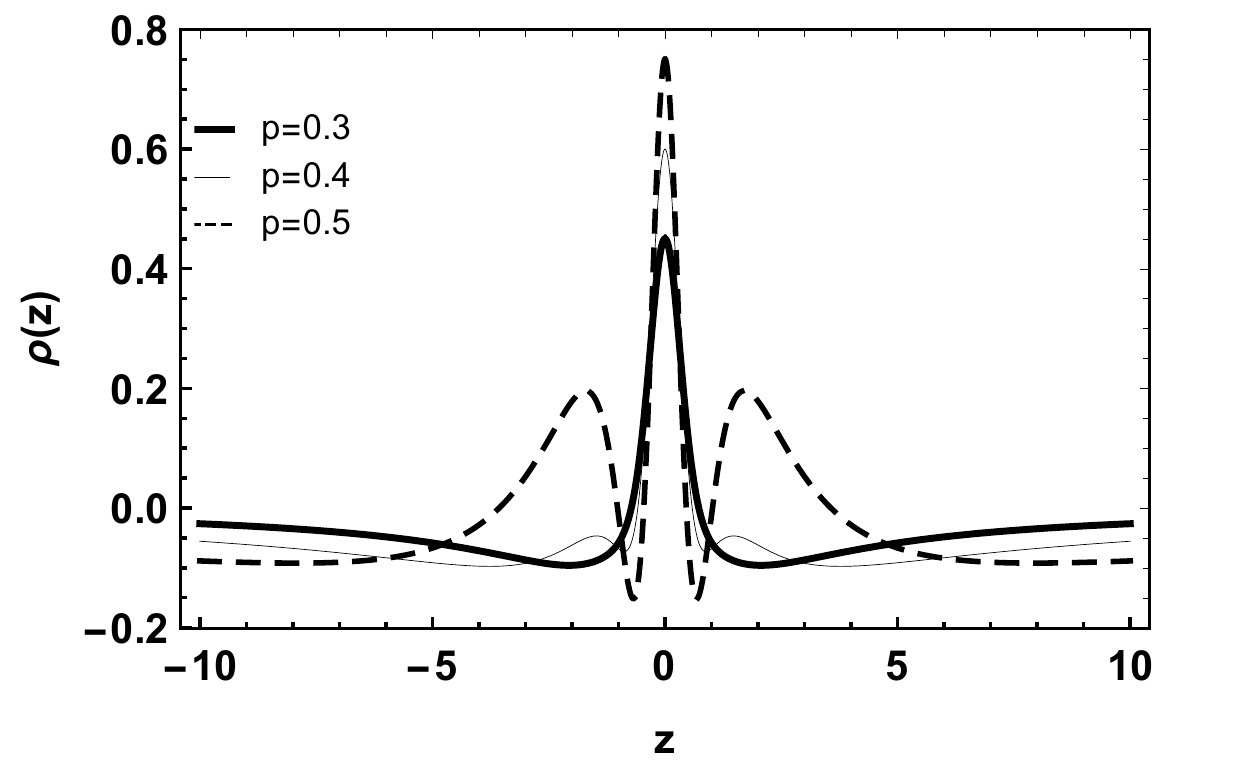}\\
(a) \hspace{8 cm}(b)\\
\includegraphics[height=5cm]{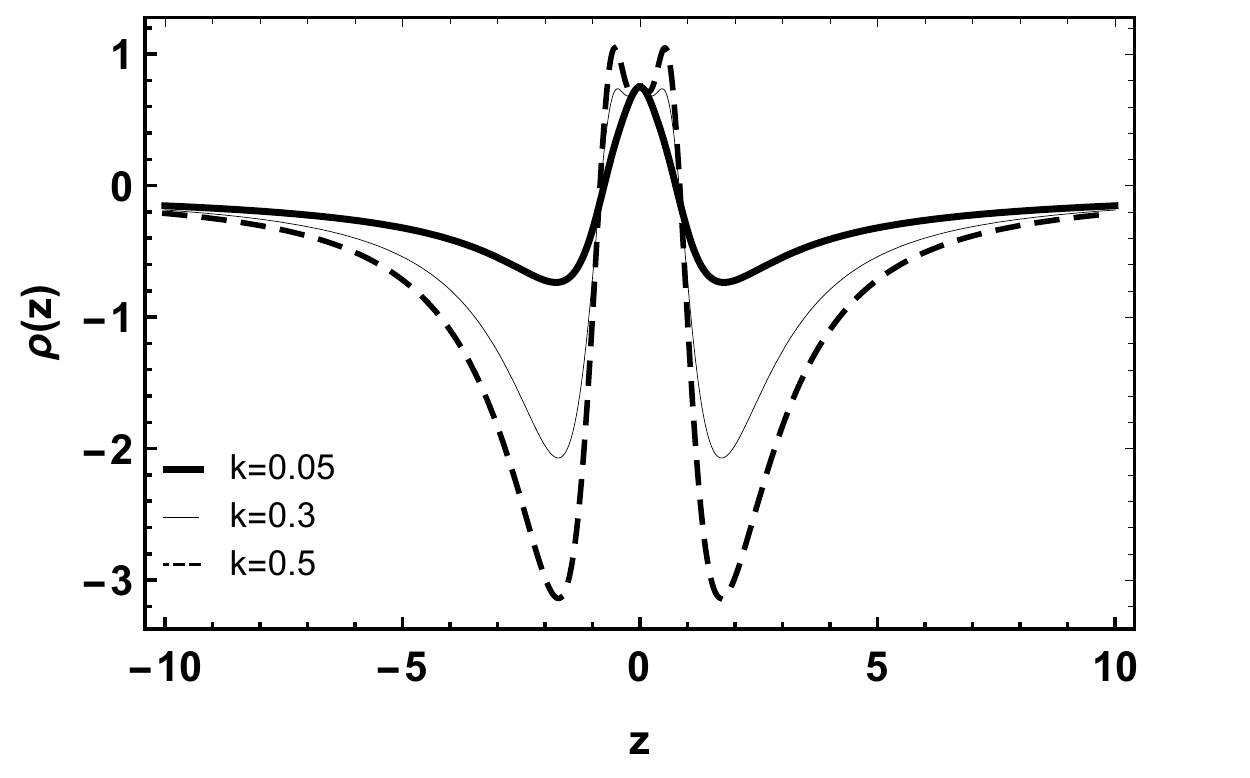} 
\includegraphics[height=5cm]{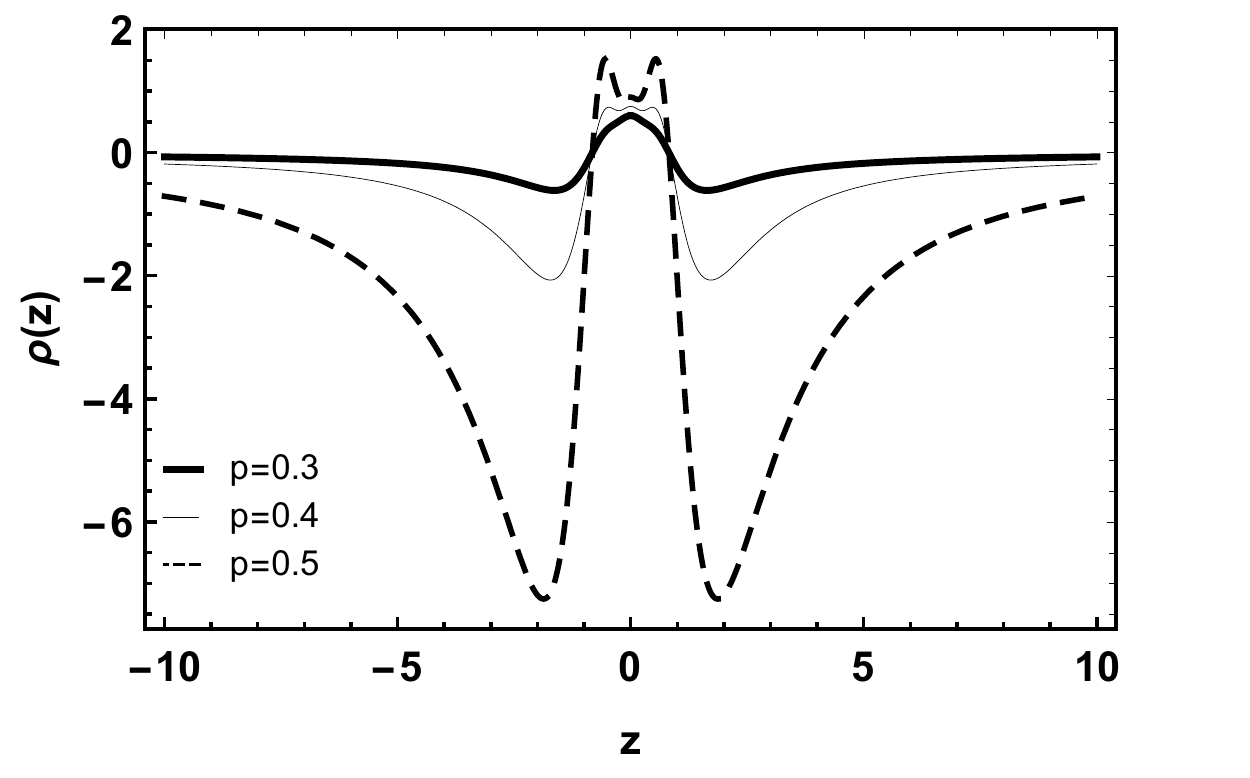}\\
(c) \hspace{8 cm}(d)
\end{tabular}
\end{center}
\caption{The shape of the energy density for $f_1$ with $\lambda=1$. (a)  $n_1=2$ and $p=0.4$. (b) $n_1=2$ and $k=0.3$. (c)  $n_1=3$ and $p=0.5$. (b) $n_1=3$ and $k=0.3$. 
\label{fig1}}
\end{figure}

We started to notice the formation of two additional peaks for the energy density starting from $n_1=2$, as we can see through the figures \ref{fig1}($a$) and \ref{fig1}($b$), and for $n_1=3$ in the figures \ref{fig1}($c$) and  \ref{fig1}($d$). This behavior is evidenced when we vary the torsion parameter $k$, and we notice that something similar happens when we vary the parameter $p$. The emergence of new peaks in energy density represents the splitting of the brane.

\subsubsection{$f_2(T)=n_2 \sinh\left(\frac{T}{n_2}\right)$}

In this second case, the energy density in the brane is
\begin{eqnarray}
\rho(z)=\frac{1}{4}\Big[n_2-\frac{\zeta^2(z^2\lambda^2-1)}{n_2 p z^2\lambda^2}\Big]\sinh\left(\frac{\zeta}{n_2}\right)-\frac{3\zeta}{8pz^2\lambda^2}\Big[(1+4p)z^2\lambda^2-1\Big]\cosh\left(\frac{\zeta}{n_2}\right).
\end{eqnarray}

\begin{figure}[ht!]
\begin{center}
\begin{tabular}{ccc}
\includegraphics[height=5cm]{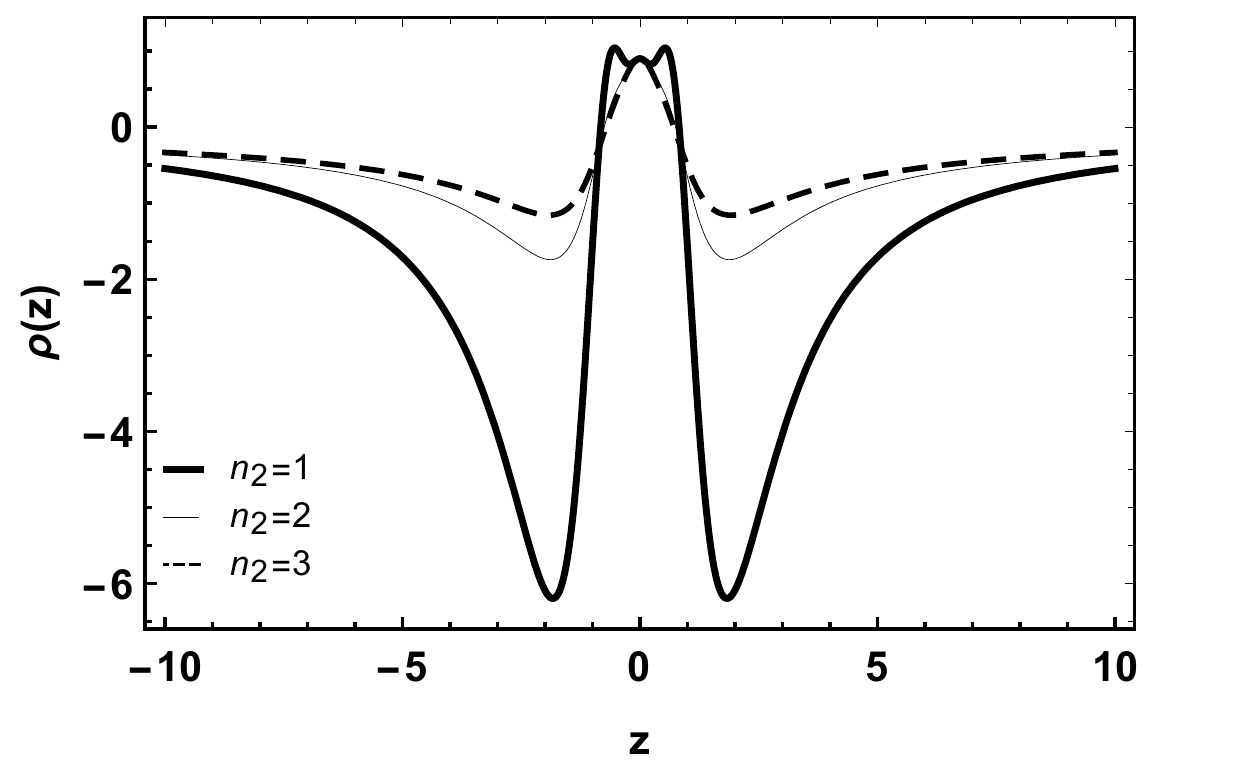} 
\includegraphics[height=5cm]{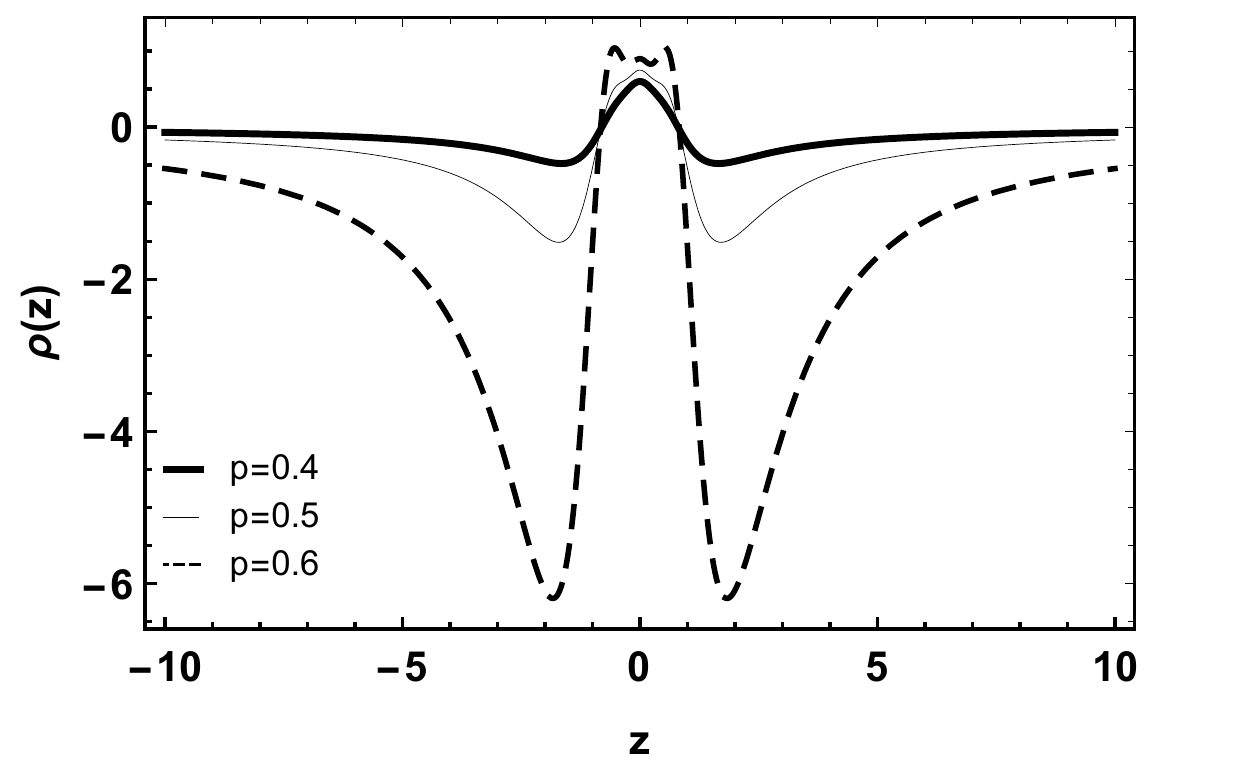}\\
(a) \hspace{8 cm}(b)
\end{tabular}
\end{center}
\caption{ The shape of the energy density for $f_2$ with $\lambda=1$. (a)   $p=0.6$. (b) $n_2=1$. 
\label{fig2}}
\end{figure}

As we can see in Fig.\ref{fig2}($a$), when we increase the value of the parameter $n_2$, new peaks appear around the core, which represents the splitting of the brane. Similar behavior happens when we vary the parameter $p$ (Fig.\ref{fig2}$b$). 

\subsubsection{$f_3(T)=n_3 \tanh\left(\frac{T}{n_3}\right)$}

Finally, in the third case, the energy density has the form
\begin{eqnarray}
\rho(z)&=&\frac{3\zeta}{8n_3pz^2\lambda^2} \Big\{4\zeta(z^2\lambda^2-1)\tanh\left(\frac{\zeta}{n_3}\right)-n_3[(1+4p)z^2\lambda^2-1]\Big\}\mathrm{sech}^2\left(\frac{\zeta}{n_3}\right)\nonumber\\ &+&\frac{n_3}{4}\tanh\left(\frac{\zeta}{n_3}\right).
\end{eqnarray}

\begin{figure}[ht!]
\begin{center}
\begin{tabular}{ccc}
\includegraphics[height=5cm]{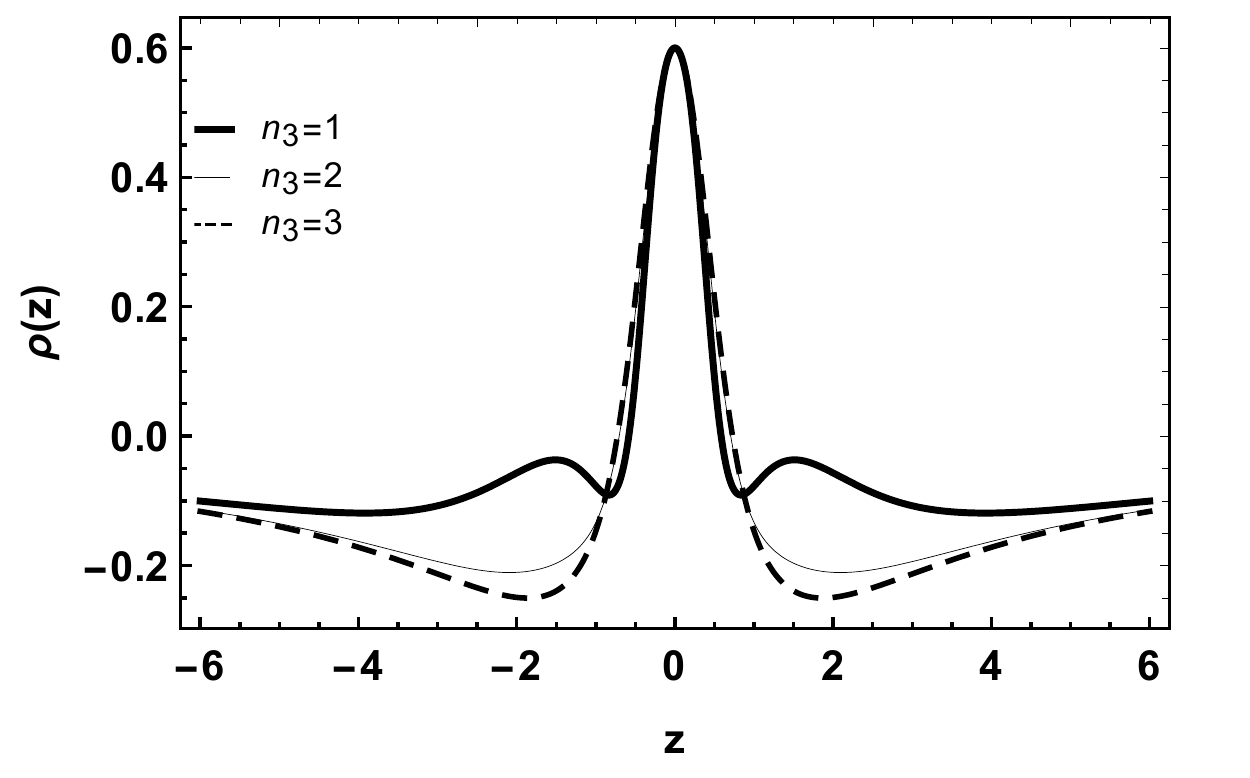} 
\includegraphics[height=5cm]{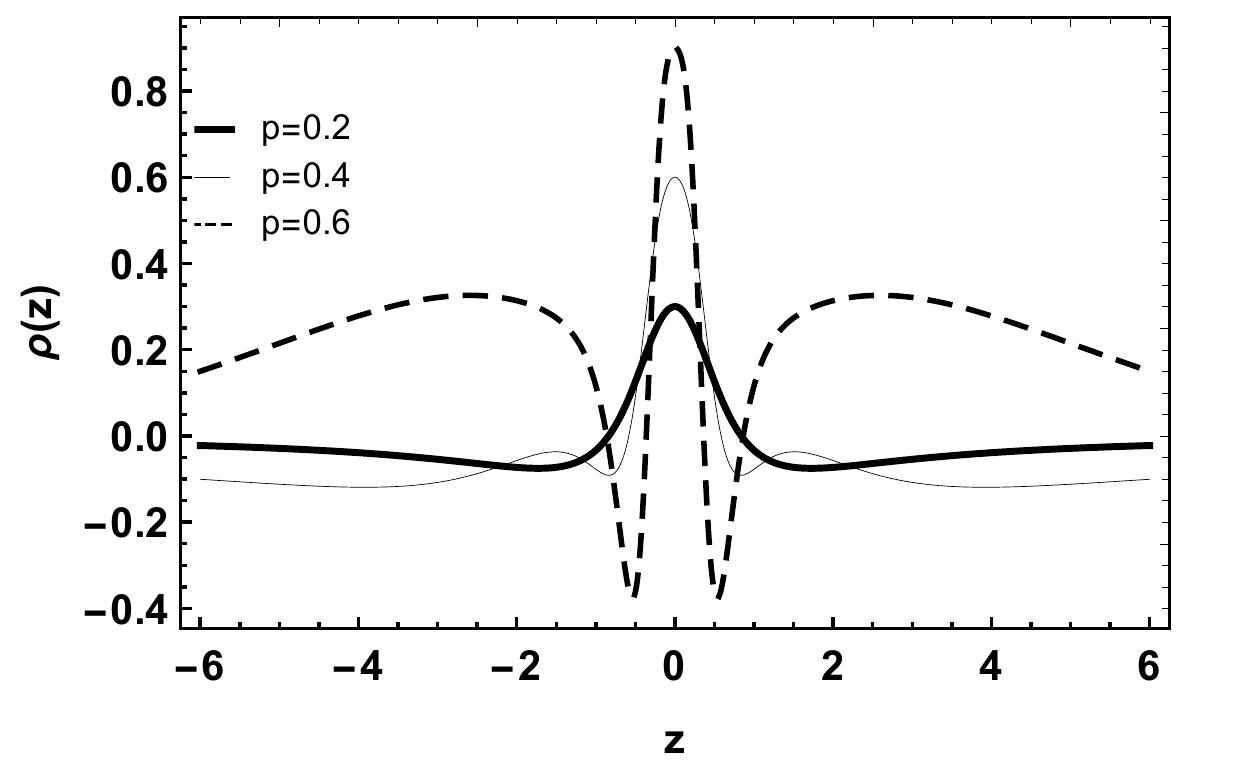}\\
(a) \hspace{8 cm}(b)
\end{tabular}
\end{center}
\caption{ The shape of the energy density for $f_3$ with $\lambda=1$. (a)   $p=0.4$. (b) $n_3=1$. 
\label{fig3}}
\end{figure}

Here results are very similar to the two previous cases $f_{1,2}(T)$. The energy densities present the splitting of the brane. This is evidenced when we increase the values of the torsion parameter $n_3$, and the same happens when we increase the value of the parameter $p$. Fig.\ref{fig3} shows the behavior of the energy density in the brane varying the parameter $n_3$ (Fig.\ref{fig3}$a$) and varying the parameter $p$ (Fig.\ref{fig3}$b$).

\section{Localization of Spin 1/2 Fermions}
\label{sec2}

The localization of fermion fields is an interesting issue. It has been known that fermions can be localized on branes in different ways. The most common coupling used in the literature to locate fermions is the coupling between fermion fields and background scalar fields. In Refs. \cite{Arias2002ew,Barbosa-Cendejas2005vog, Barbosa-Cendejas:2006cic}, a single bound state and a continuous gapless spectrum of massive fermion Kaluza-Klein (KK) states can be obtained, with different scalar-fermion couplings. On the other hand, in Refs. \cite{Liu:2008pi,Liu:2009uca,Barbosa-Cendejas:2007cwl,Liu:2009dwa,Alencar:2012en} it was shown that in some  models, there exist finite discrete KK states and a continuous gapless spectrum.

In Ref. \cite{Yang2012}, Yang and collaborators investigate how the spacetime torsion influences the localization of fermion fields on the brane, taking a Yukawa-type coupling of a massless spin $1/2$ fermion to the background scalar $\phi$. The influence of the boundary term on the location of fermion fields on the brane is also investigated using the same Yukawa-type coupling in Ref. \cite{Moreira20211}. Now, we can then raise a question: What would happen if we took a non-minimal coupling torsion-fermion? It is inspired by this questioning that we developed this section.

Taking a non-minimal coupling with $g(T)$, the Dirac action of a massless spin $1/2$ fermion in five dimensions can be written as
\begin{eqnarray}\label{1}
\mathcal{S}_{1/2}=\int h \overline{\Psi} \Big(\Gamma^M D_M\Psi -\xi g(T)\Psi\Big)d^5x,
\end{eqnarray}
being $\Gamma^M=h_{\overline{M}}\ ^M \Gamma^{\overline{M}}$  the Dirac curved matrices ($\Gamma^{\overline{M}}$ are the Dirac flat matrices), and $ D_M=\partial_M +\Omega_M$ the covariant derivative, where the spin connection is given by \cite{Obukhov2002, Ulhoa2016,Moreira20211}
\begin{eqnarray}\label{3}
\Omega_M=\frac{1}{4}\Big(K_M\ ^{{\overline{N}}{\overline{Q}}}\Big)\ \Gamma_{\overline{N}}\Gamma_{\overline{Q}}.
\end{eqnarray}
For our case, the spinor representation is \cite{Almeida2009, Dantas, Andrade2001} 
\begin{eqnarray}
\Psi\equiv\Psi(x,z)=\left(\begin{array}{cccccc}
\psi\\
0\\
\end{array}\right),\ 
\Gamma^{\overline{\mu}}=\left(\begin{array}{cccccc}
0&\gamma^{\overline{\mu}}\\
\gamma^{\overline{\mu}}&0\\
\end{array}\right),\ \Gamma^{\overline{z}}=\left(\begin{array}{cccccc}
0&\gamma^4\\
\gamma^4&0\\
\end{array}\right),
\end{eqnarray}
and the spin connections are $\Omega_\mu=\frac{1}{4}(-\partial_z A)\Gamma_{\mu}\Gamma^{z}$ and $\Omega_z=\partial_z A$. So, the Dirac equation takes the form
\begin{eqnarray}\label{7}
\Big[\gamma^{\mu}\partial_\mu+\gamma^4\partial_z-\xi e^Ag(T)\Big]\psi=0.
\end{eqnarray}
We can decompose Eq.(\ref{7}) applying the properties $\psi=\sum_n[\psi_{L,n}(x)\varphi_{L,n}(z)+\psi_{R,n}(x)\varphi_{R,n}(z)]$, being $\gamma^4\psi_{R,L}=\pm\psi_{R,L}$ and $\gamma^\mu\partial_\mu\psi_{R,L}=m\psi_{L,R}$, which gives us two coupled equations
\begin{eqnarray}\label{9}
\Big[\partial_z+\xi e^A g(T)\Big]\varphi_{L}(z)=m \varphi_{R}(z),\nonumber\\
\Big[\partial_z-\xi e^Ag(T)\Big]\varphi_{R}(z)=m \varphi_{L}(z).
\end{eqnarray}
Eqs.(\ref{9}) are easily decoupled, generating Schroëdinger-like equations
\begin{eqnarray}\label{10}
\Big[-\partial^2_z+V_L(z)\Big]\varphi_{L}(z)=m^2 \varphi_{L}(z),\nonumber\\
\Big[-\partial^2_z+V_R(z)\Big]\varphi_{R}(z)=m^2 \varphi_{R}(z),
\end{eqnarray}
where
\begin{eqnarray}\label{11}
V_L(z)=U^2 -\partial_{z}U,\nonumber\\
V_R(z)=U^2 +\partial_{z}U,
\end{eqnarray}
and $U=\xi e^A g(T)$ is the so-called superpotential.

\subsection{Massless fermionic modes}

Likewise in the just mentioned Yukawa-type coupling, the form of the non-minimal coupling $g(T)$ is fundamental for the dynamics of the KK states.
Massless modes (zero modes) have the form
\begin{eqnarray}
\varphi_{R0,L0}(z)\propto \exp{\Bigg[\pm\int\xi g(T)e^{A}dz\Bigg]},
\end{eqnarray}
due to the supersymmetric structure of the potentials \eqref{11}. 

We need to impose some conditions in the form of $g(T)$. Based on the Yukawa-type coupling ($\overline{\Psi} \phi\Psi$), which is the
common in the literature, we can propose the following conditions for $g(T)$:
The function $g(T(z))$ must be antisymmetric when undergoing a phase transition at the origin $g(T(z=0))=0$ (similar to the solution of $\phi$). Also, $g(T(z\rightarrow\pm\infty))$ must go to some constant value $c$ asymptotically. For our model to make physical sense, the so-called superpotential $U(g(T))|_{z\rightarrow\pm\infty}\rightarrow 0$ must go to zero in vacuum.

To check whether the zero modes can be localized on the brane, we should check whether the normalization condition for the zero modes is satisfied, or in other words, whether the integral is satisfied
\begin{eqnarray}
\int \varphi_{R0,L0}(z)^2dz<\infty.
\end{eqnarray}
Since $g(T)e^{A}|_{z\rightarrow\pm\infty}\rightarrow0$, only the left-handed massless mode is trapped in the brane (for positive $\xi$), a feature shared by the Yukawa coupling models \cite{Yang2012}.

Since $p=1$, we can get a simple expression of the warp factor in conformal coordinate, i.e., $A(z)=-\ln\sqrt{1+\lambda^2z^2}$. At the infinity, $e^A\rightarrow1/\lambda|z|$, hereby,
\begin{eqnarray}
\varphi_{L0}(z\rightarrow\pm\infty)\rightarrow|z|^{-\xi g(T)_{\infty}/\lambda},
\end{eqnarray}
where $g(T)_{\infty}=g(T(z\rightarrow\infty))=c$. If the normalization condition is satisfied, we can get the following equivalent condition,
\begin{eqnarray}
\int |z|^{-2\xi g(T)_{\infty}/\lambda}dz<\infty.
\end{eqnarray}
Only when $\xi>\lambda/2g(T)_{\infty}$, the above integral is convergent,
which means that the left-chiral zero modes can be localized on the brane under this condition.

Based on these conditions, we propose a model of $g(T)$ for each case of $f(T)$ discussed in the previous section. Our choices of $g(T)$ were based on providing greater mathematical simplicity.
For the first case $f_1(T)$, we can propose
\begin{eqnarray}
g_1(T)=\sqrt{-f_1(T)},
\end{eqnarray}
where the so-called superpotential takes the form
\begin{eqnarray}
U(z)=\frac{\xi\sqrt{\zeta}}{(1+z^2\lambda^2)^{\frac{p}{2}}}\sqrt{1+k\zeta^{n_1-1}},
\end{eqnarray}
remembering that $\zeta= 12p^2z^2\lambda^4(1+z^2\lambda^2)^{p-2}$.

\begin{figure}[ht!]
\begin{center}
\begin{tabular}{ccc}
\includegraphics[height=5cm]{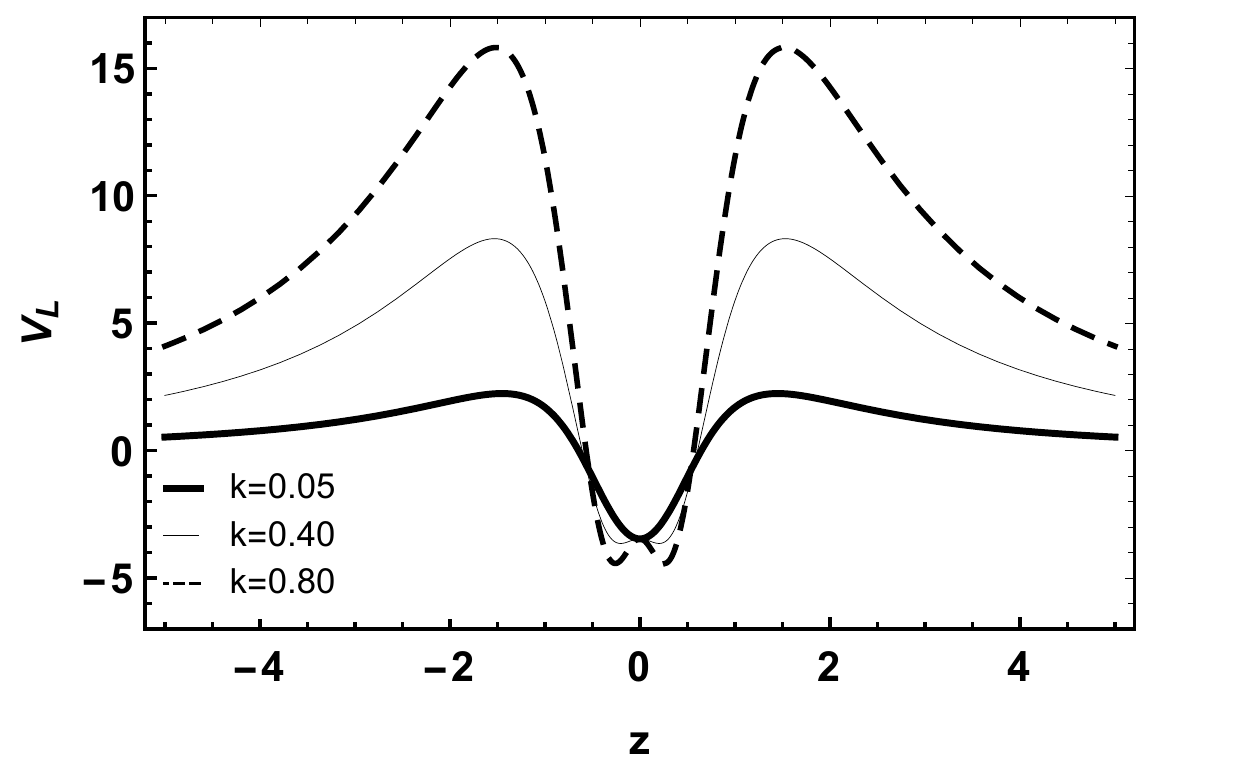} 
\includegraphics[height=5cm]{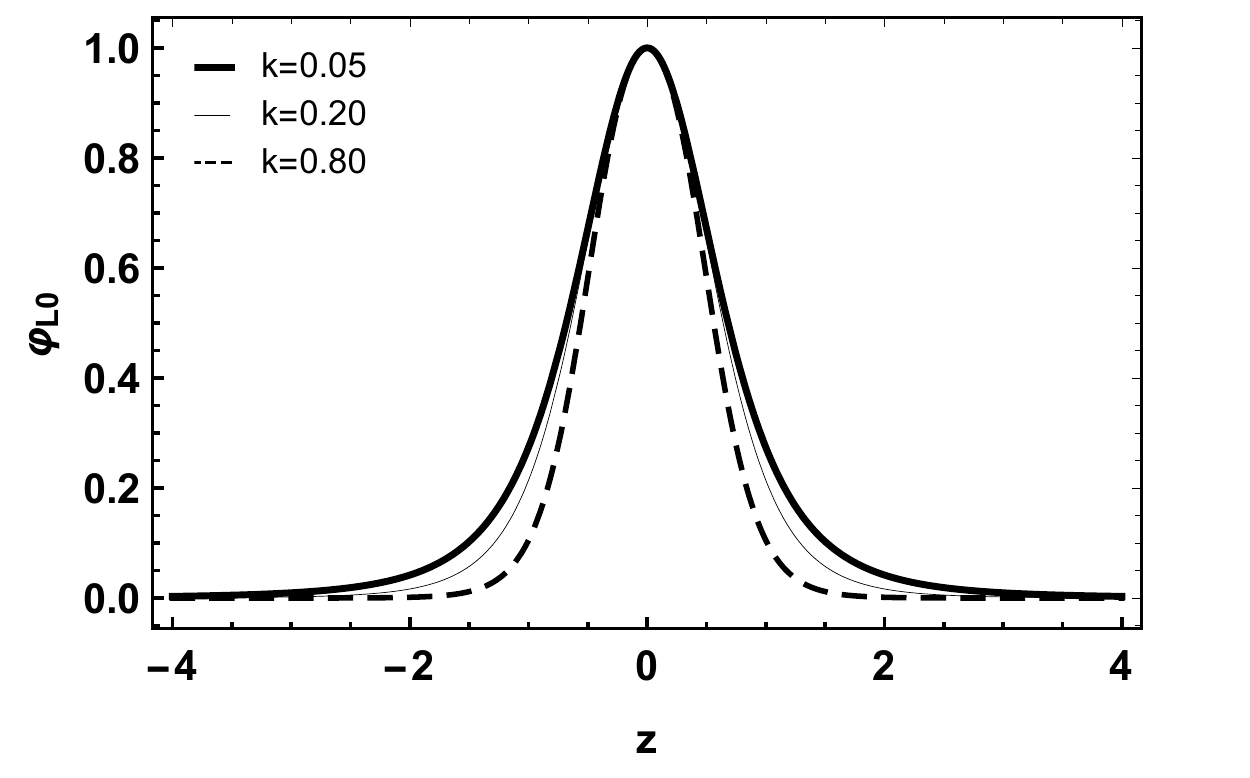}\\
(a) \hspace{8 cm}(b)\\
\includegraphics[height=5cm]{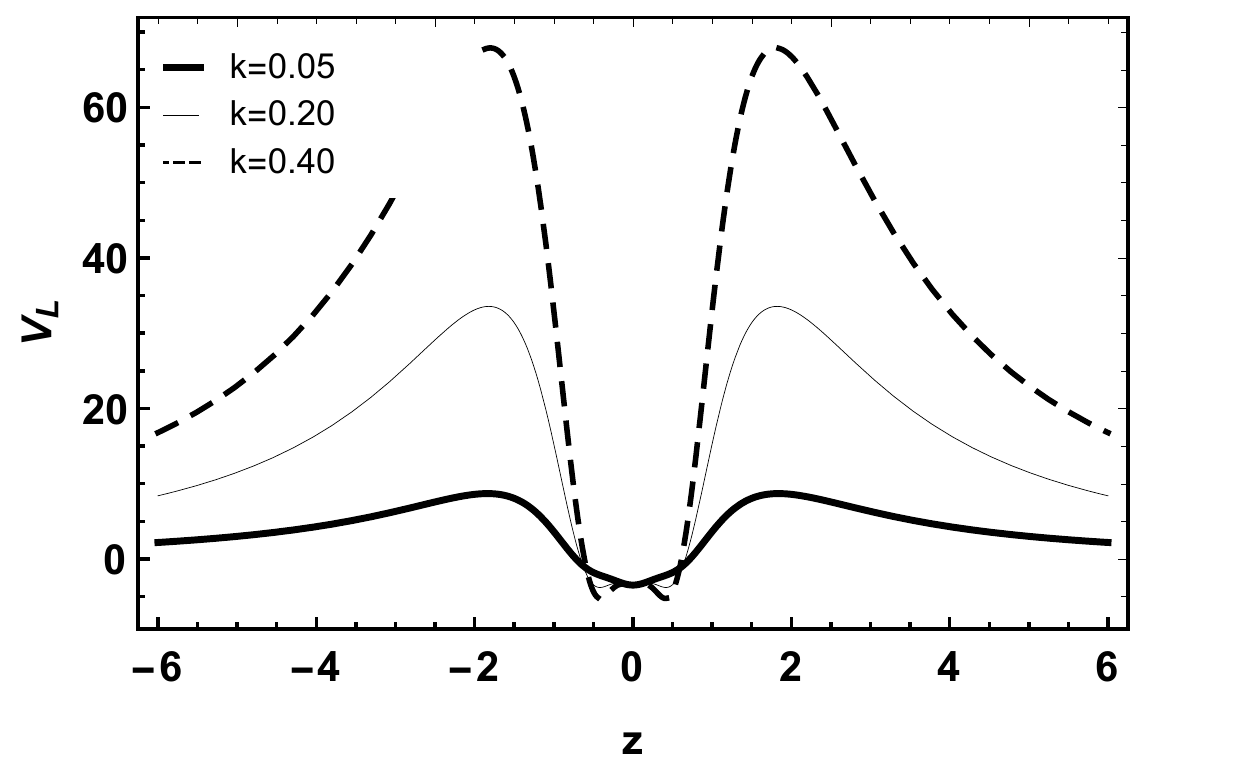} 
\includegraphics[height=5cm]{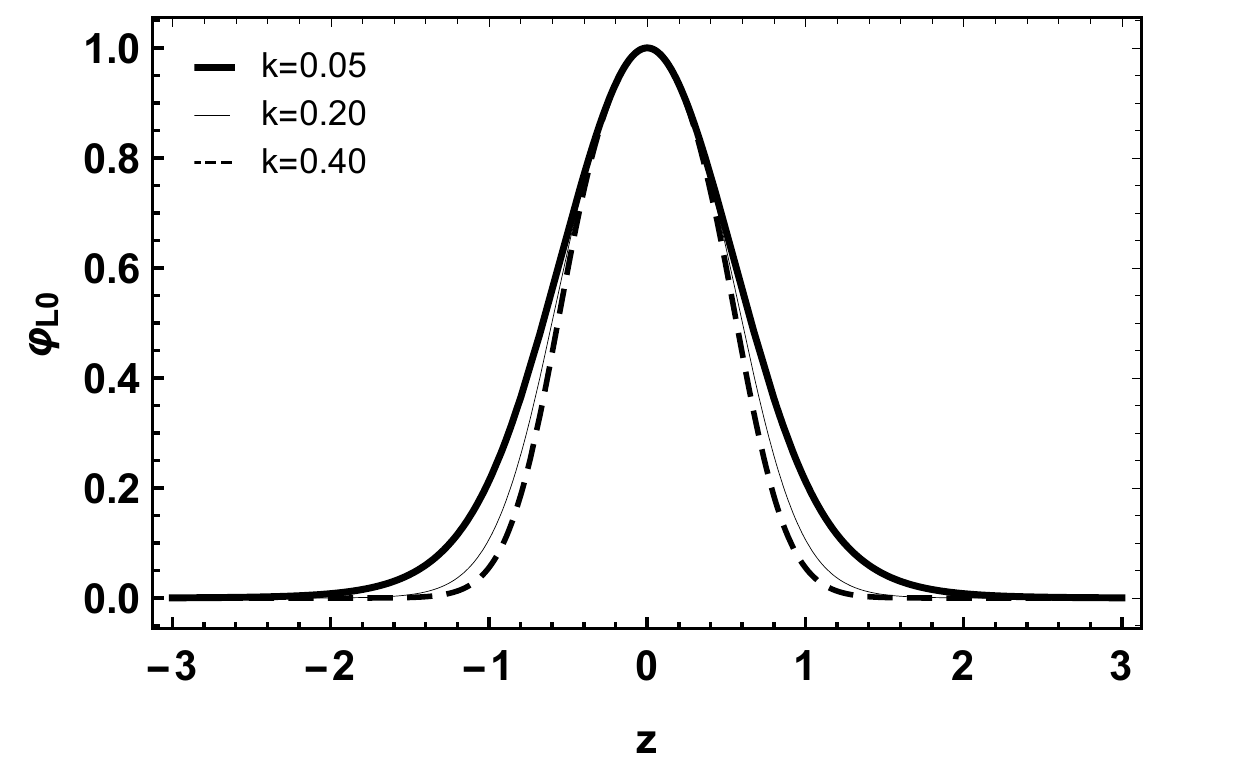}\\
(c) \hspace{8 cm}(d)
\end{tabular}
\end{center}
\caption{ For $\lambda=p=\xi=1$ and $g_1$.  For $n_1=2$, (a) 
Potential $V_L$ and (b) Massless mode $\varphi_{L0}$. For $n_1=3$, (c) 
Potential $V_L$ and  d) Massless mode $\varphi_{L0}$.
\label{fig4}}
\end{figure}

The expression of the potential $V_L(z)$ is quite large, so for simplicity, we show its behavior through Figs.\ref{fig4}($a$) and \ref{fig4}($c$). We note that for $n_1=2$ (Fig.\ref{fig4}$a$), the torsion parameter $k$ directly affects the behavior of the potential, intensifying the two peaks away from the core and causing the appearance of two new wells around the core.  The same happens with $n_1=3$ (Fig.\ref{fig4}$c$).

Massless fermionic modes are obtained numerically. Their behavior is shown in Figs. \ref{fig4}($b$) for $n_1=2$ and \ref{fig4}($d$) for $n_1=3$. We can see that massless modes are also affected by the torsion parameter $k$, becoming more localized as the parameter $k$ increases.

For the second case $f_2(T)$, we do
\begin{eqnarray}
g_2(T)=f_2(\sqrt{-T}),
\end{eqnarray}
where the so-called superpotential takes the form
\begin{eqnarray}
U(z)=\frac{\xi n_2}{(1+z^2\lambda^2)^{\frac{p}{2}}}\sinh\left(\frac{\sqrt{\zeta}}{n_2}\right).
\end{eqnarray}

\begin{figure}[ht!]
\begin{center}
\begin{tabular}{ccc}
\includegraphics[height=5cm]{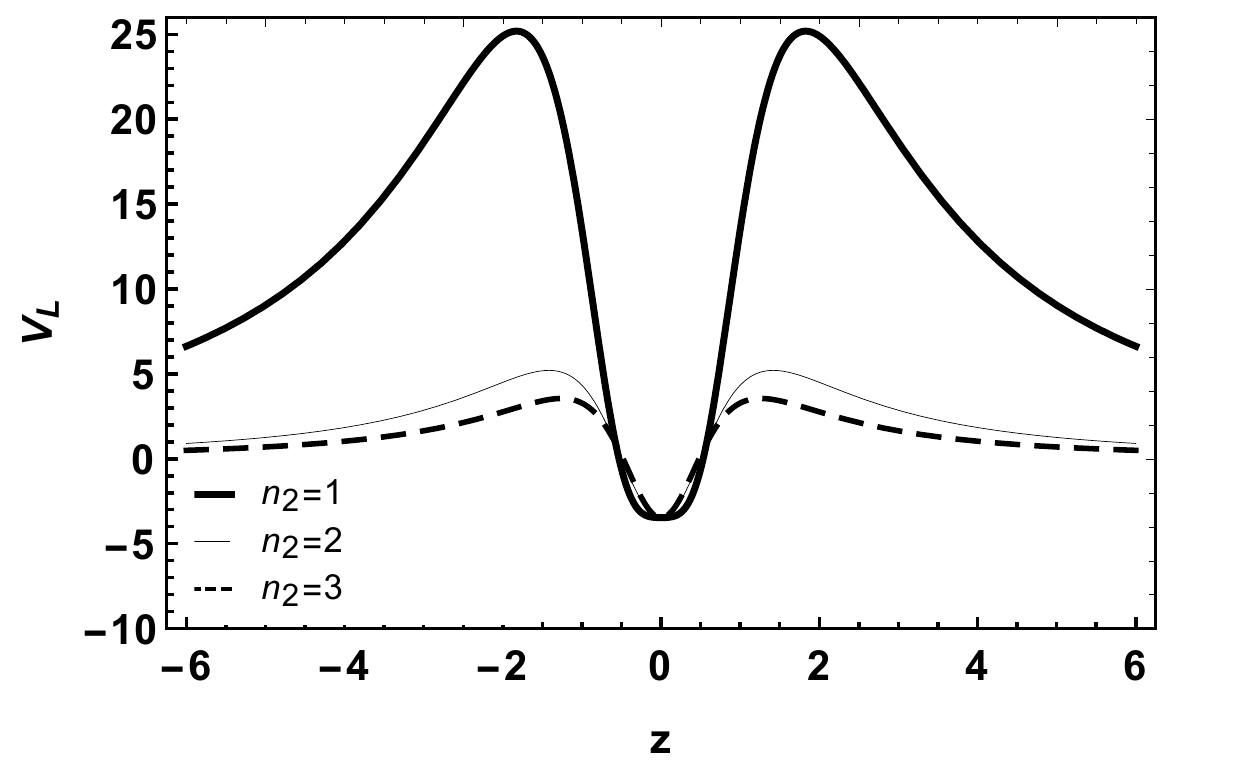} 
\includegraphics[height=5cm]{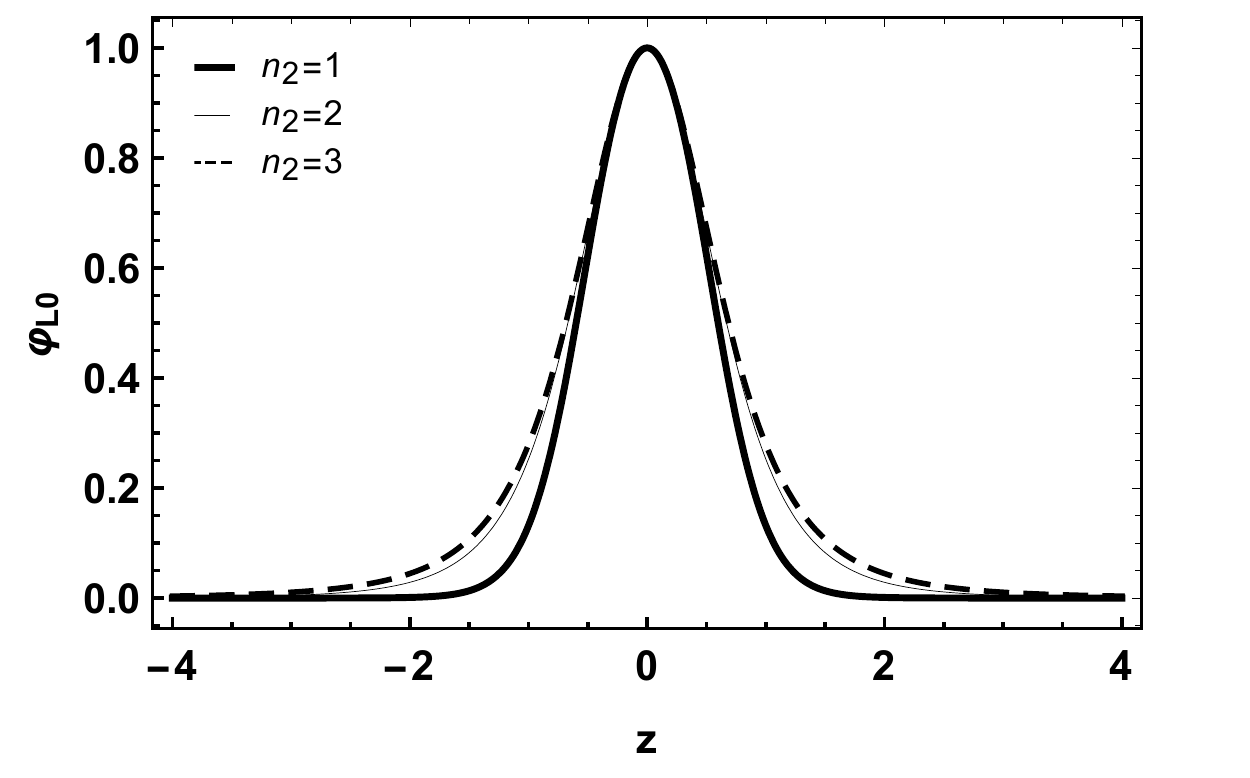}\\
(a) \hspace{8 cm}(b)
\end{tabular}
\end{center}
\caption{ Being $\lambda=p=\xi=1$ for $g_2$. (a) Potential $V_L$. (b) 
Massless mode $\varphi_{L0}$. 
\label{fig5}}
\end{figure}

Fig.\ref{fig5}($a$) shows the behavior of the potential $V_L(z)$.  We note that the torsion parameter $n_2$ directly affects the behavior of the potential. When we decrease the value of $n_2$, it intensifies the two peaks away from the core and causes a decrease in the width of the well. This behavior affects massless modes, which become more localized when we decrease the value of the parameter $n_2$ (Fig. \ref{fig5}$b$).

Finally, for the third case $f_3(T)$, we do the same as in the case $f_2(T)$, where
\begin{eqnarray}
g_3(T)=f_3(\sqrt{-T}),
\end{eqnarray}
so, the so-called superpotential takes the form
\begin{eqnarray}
U(z)=\frac{\xi n_3}{(1+z^2\lambda^2)^{\frac{p}{2}}}\tanh\left(\frac{\sqrt{\zeta}}{n_3}\right).
\end{eqnarray}

\begin{figure}[ht!]
\begin{center}
\begin{tabular}{ccc}
\includegraphics[height=5cm]{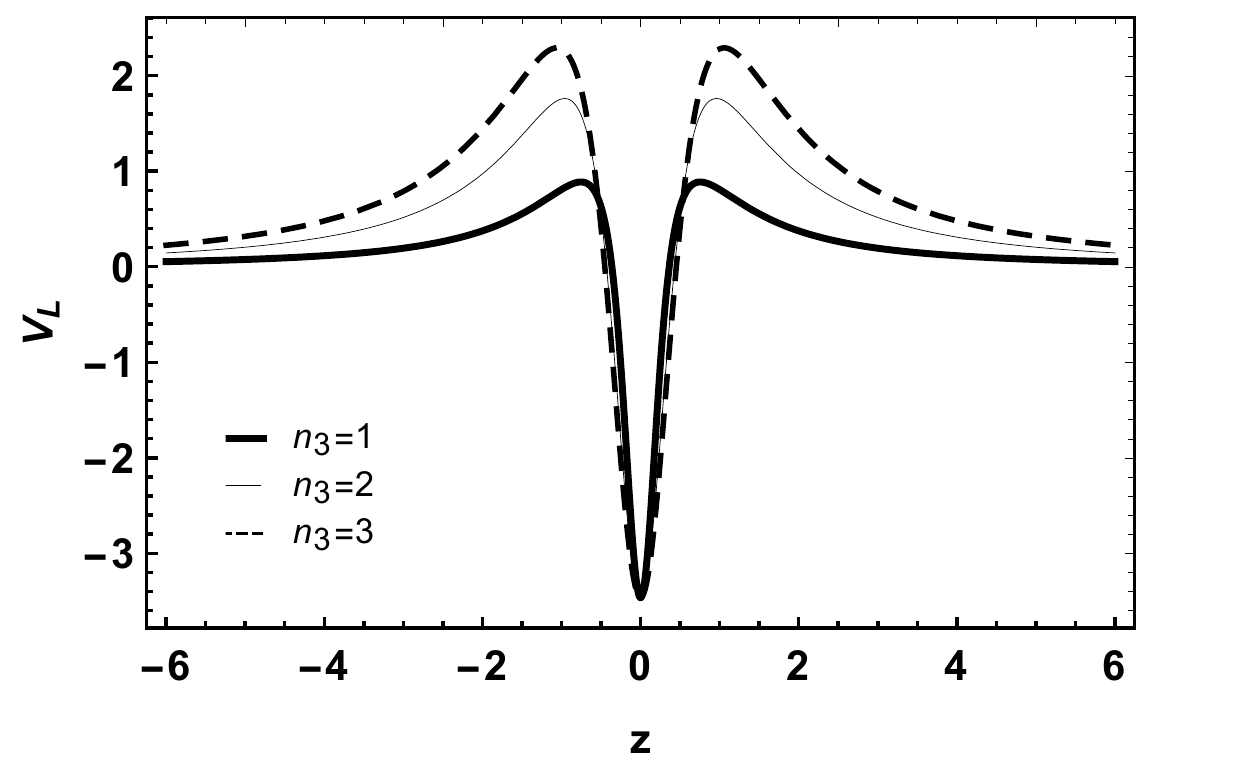} 
\includegraphics[height=5cm]{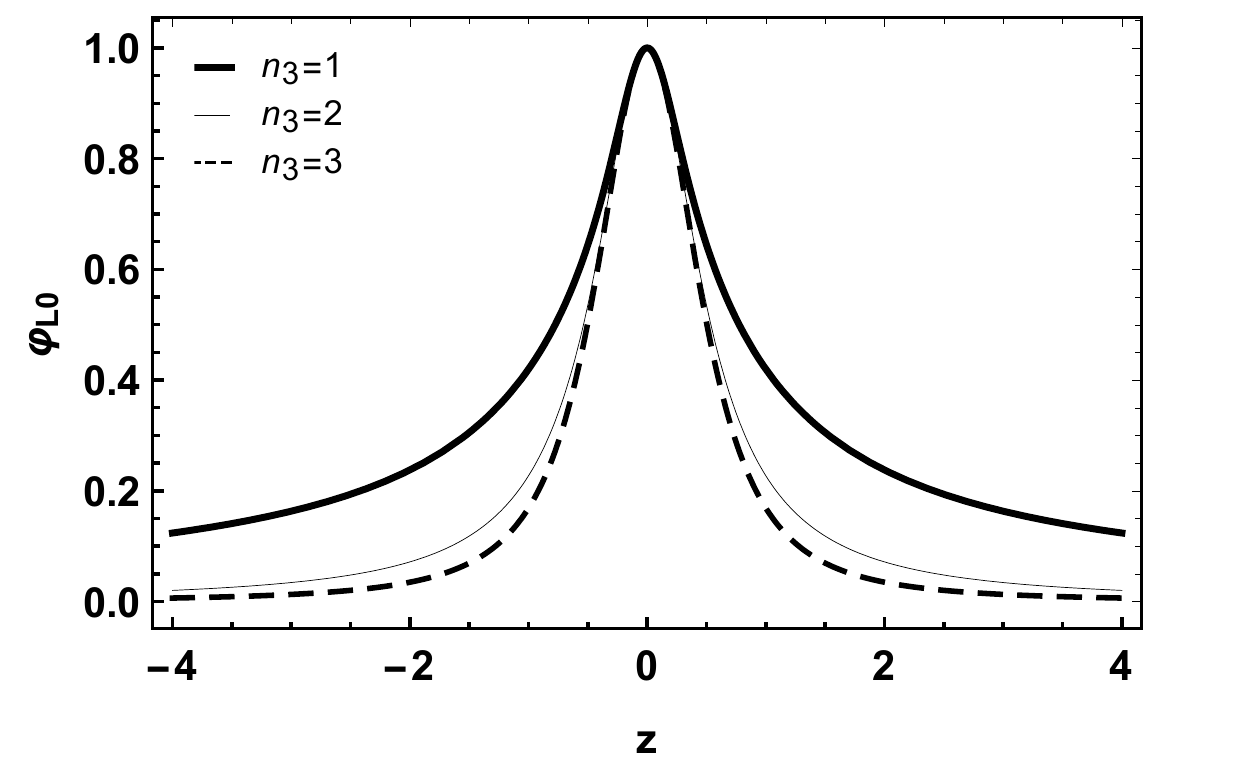}\\
(a) \hspace{8 cm}(b)
\end{tabular}
\end{center}
\caption{ Being $\lambda=p=\xi=1$ for $g_3$. (a) Potential $V_L$. (b) Massless mode $\varphi_{L0}$.
\label{fig6}}
\end{figure}

Fig.\ref{fig6}($a$) shows the behavior of the potential $V_L(z)$.  We note that the torsion parameter $n_3$ directly affects the behavior of the potential, but, contrary to the $g_2$ case, were increasing the value of $n_3$, intensifying the two peaks away from the core and causing the increase in the width of the well. This behavior affects massless modes, which become more localized when we increase the value of the parameter $n_3$ (Fig.\ref{fig6}$b$).

\subsection{Massive fermionic mode}

Knowing the behavior of the effective potentials, represented in Figs. \ref{fig4}$a$,  \ref{fig4}$c$. \ref{fig5}$a$, and \ref{fig6}$a$, we can see that they are even functions. Then we can find the massive fermionic modes by numerically solving Eqs. (\ref{10}), keeping in mind that the wave functions will be even or odd. Therefore, the boundary conditions are
\begin{eqnarray}
\varphi_{even}(0)&=&c,\ \ \partial_z\varphi_{even}(0)=0,\nonumber\\
\varphi_{odd}(0)&=&0, \ \ \partial_z\varphi_{odd}(0)=c,
\end{eqnarray}
where $c$ is just a constant \cite{Almeida2009,Liu2009,Liu2009a}. It is important to note that here $\varphi_{even}$ and $\varphi_{odd}$ respectively represent the even and odd parity modes of $\varphi_{R,L}(z)$.

Although the massive fermionic mode is not located on the brane, some massive states can exhibit a relatively large amplitude close to the brane \cite{Almeida2009}.
These massive states occur for potentials that present a potential well close to the brane and for masses $m^2$ up to the maximum value of the potential barrier \cite{Liu2009,Liu2009a}. We define these states as resonant modes, and they can be obtained by the quantum mechanical analog structure of massive modes \cite{Almeida2009,Liu2009,Liu2009a}.

To identify the solutions of the Schrodinger-like equation (\ref{10}) with the largest amplitudes close to the brane, we use the resonance method. The relative probability $P(m)$ of finding a particle with mass $m$ in a narrow band $2z_b$ is \cite{Almeida2009,Liu2009,Liu2009a}
\begin{eqnarray}
P_{R,L}(m)=\frac{\int_{-z_b}^{z_b} |\varphi_{R,L}(z)|^2 dz}{\int_{-z_{max}}^{z_{max}} |\varphi_{R,L}(z)|^2 dz},
\end{eqnarray}
where $z_{max}$ stands to the domain limits. Larger values of the parameter
$z_b$ do not change the results for the position of the resonance peaks, but small values of $z_b$ are more efficient to identify the peaks.

\begin{figure}[ht!]
\begin{center}
\begin{tabular}{ccc}
\includegraphics[height=5cm]{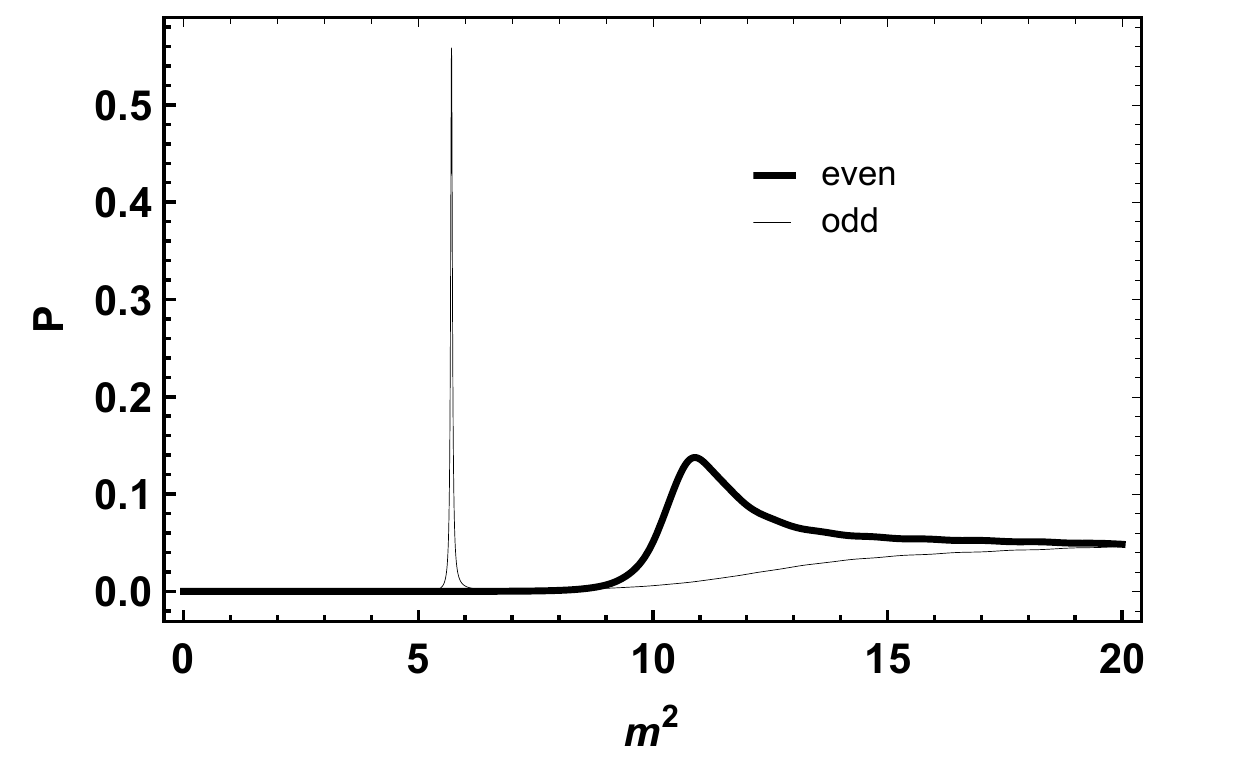}
\includegraphics[height=5cm]{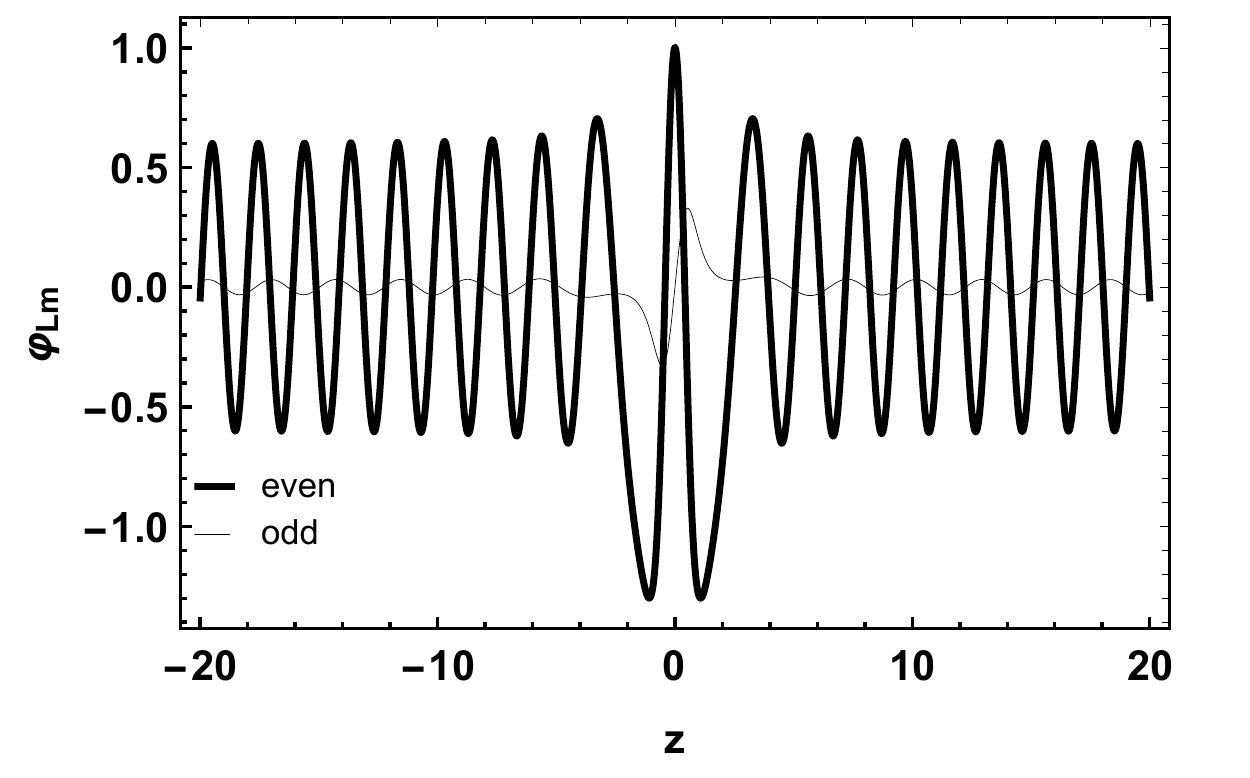}\\
(a)\hspace{6 cm}(b)\\
\includegraphics[height=5cm]{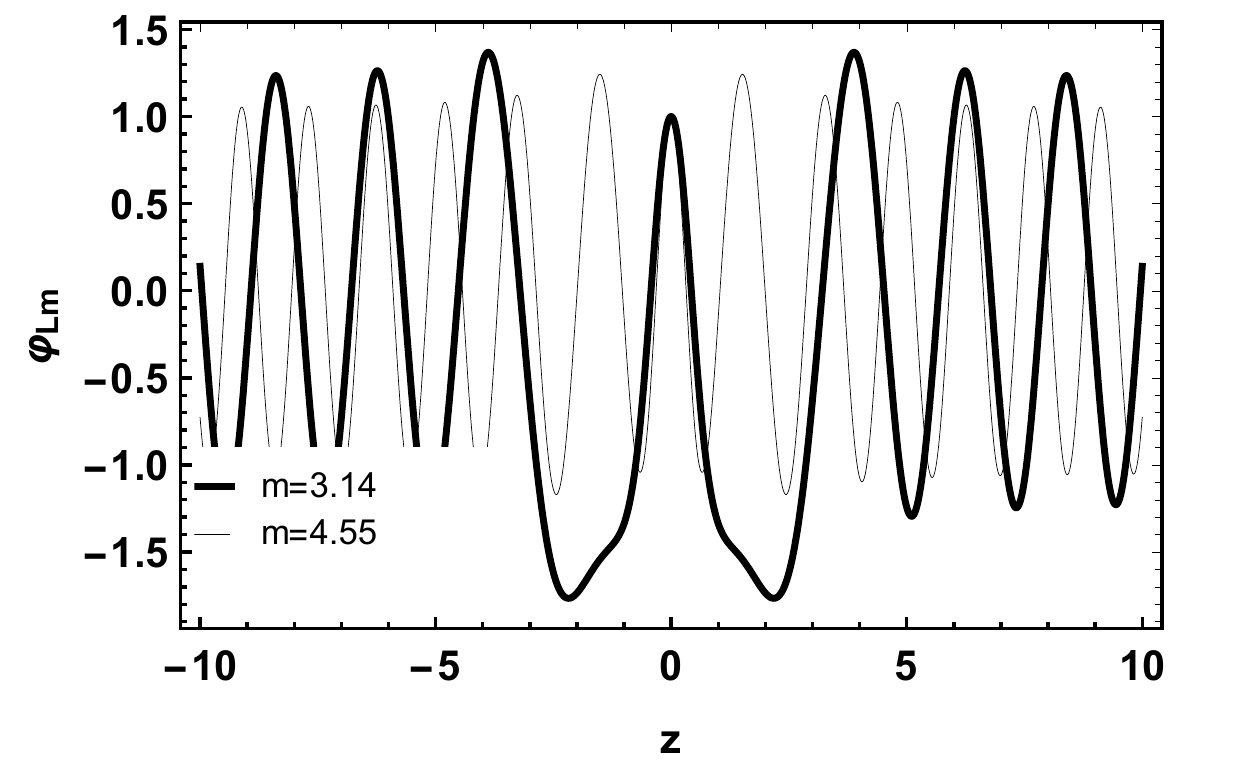} 
\includegraphics[height=5cm]{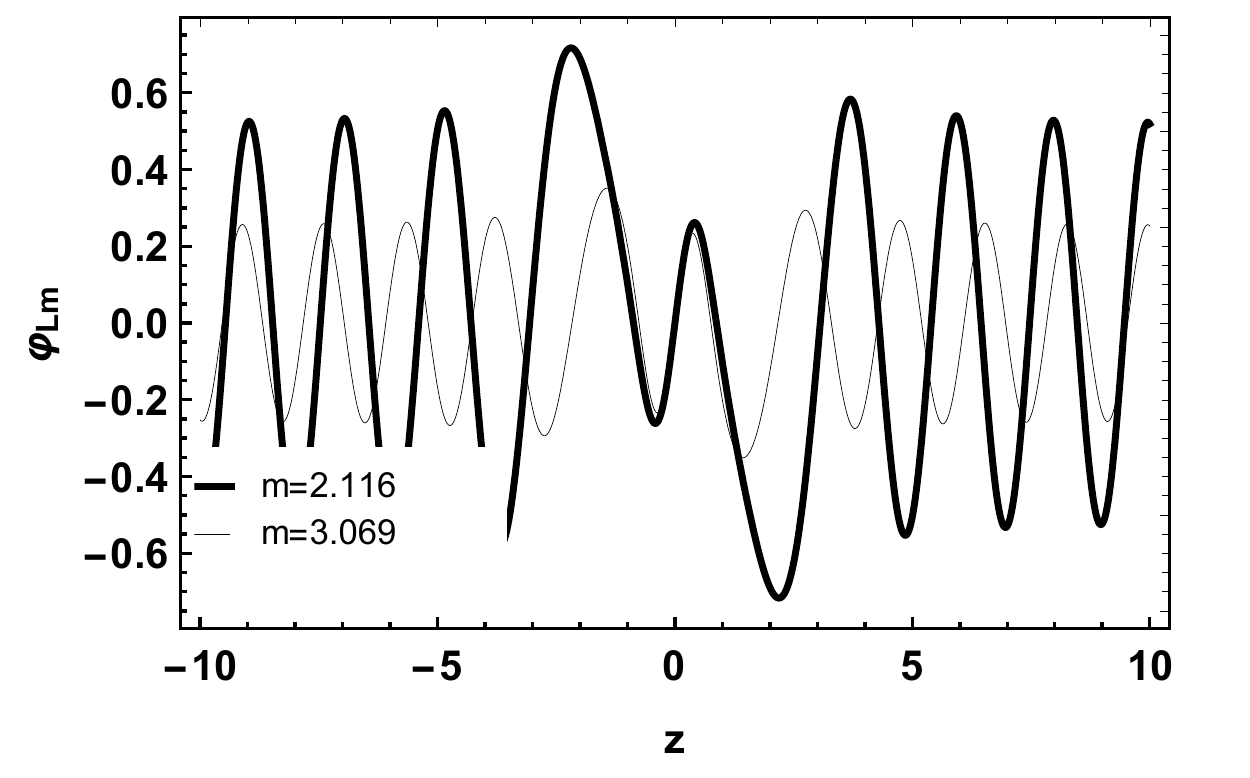}\\
(c) \hspace{6 cm}(d)\\
\includegraphics[height=5cm]{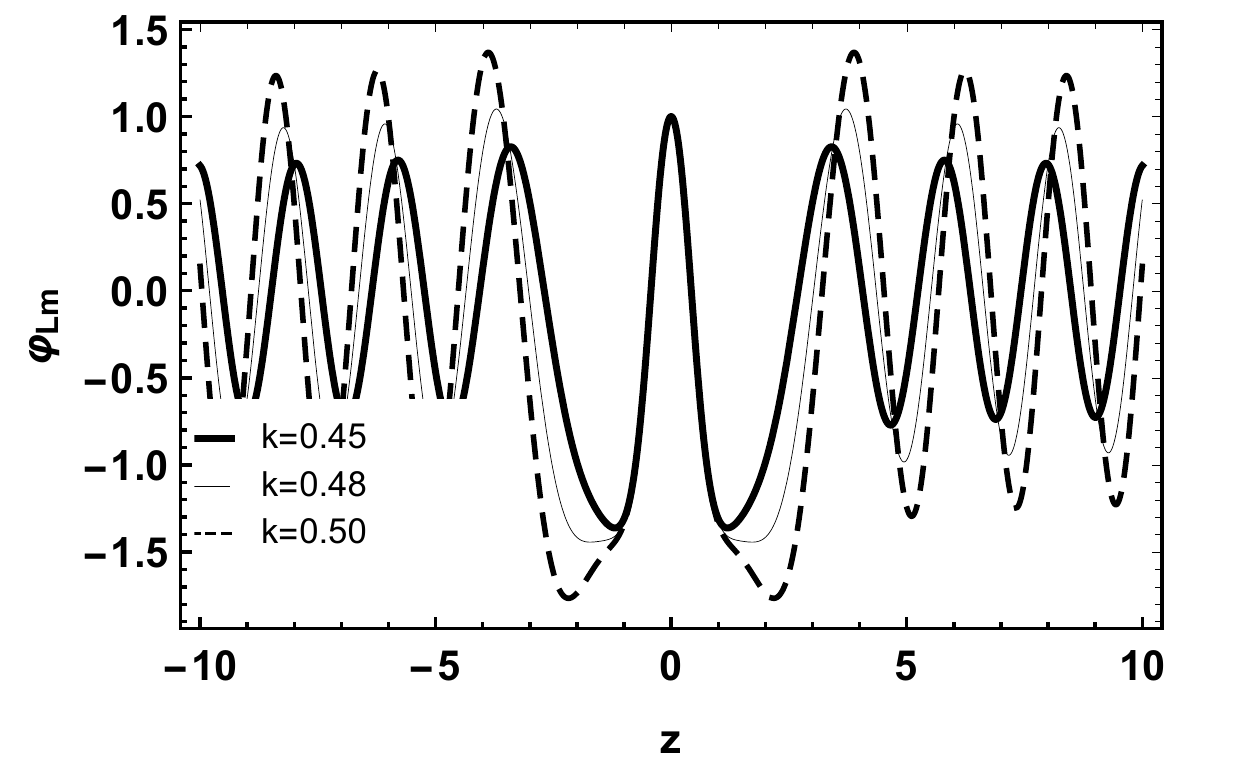} 
\includegraphics[height=5cm]{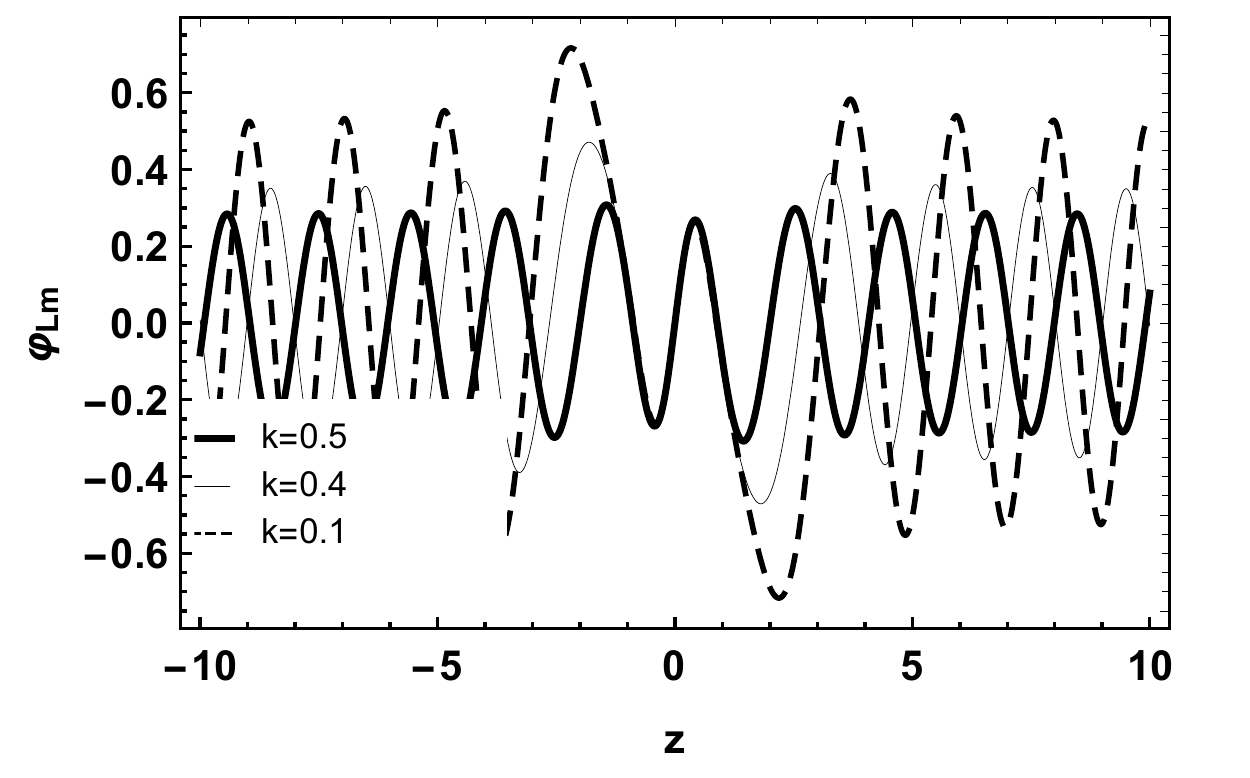}\\
(e) \hspace{6 cm}(f)
\end{tabular}
\end{center}
\caption{For $g_1(T)$, with $n_1=2$, $k=0.5$ and $p=\lambda=\xi=1$. (a) The form of the relative probability $P_L(m)$. (b) The resonant fermionic modes. The massive fermionic modes for $\varphi_{even}$ (c) and $\varphi_{odd}$ (d). By varying $k$, $\varphi_{even}$ with $m=3.14$ (e) and $\varphi_{odd}$ with $m=2.116$ (f).}
\label{fig7.1}
\end{figure}

\begin{figure}[ht!]
\begin{center}
\begin{tabular}{ccc}
\includegraphics[height=5cm]{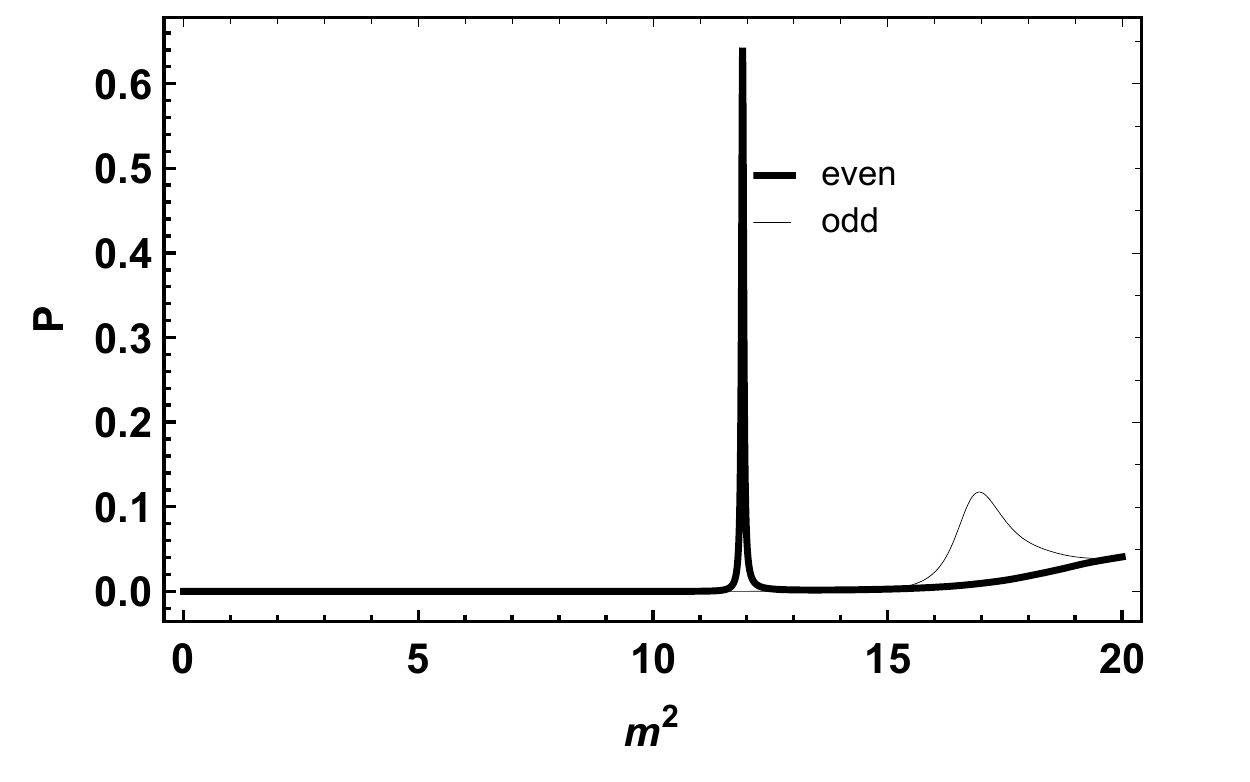}
\includegraphics[height=5cm]{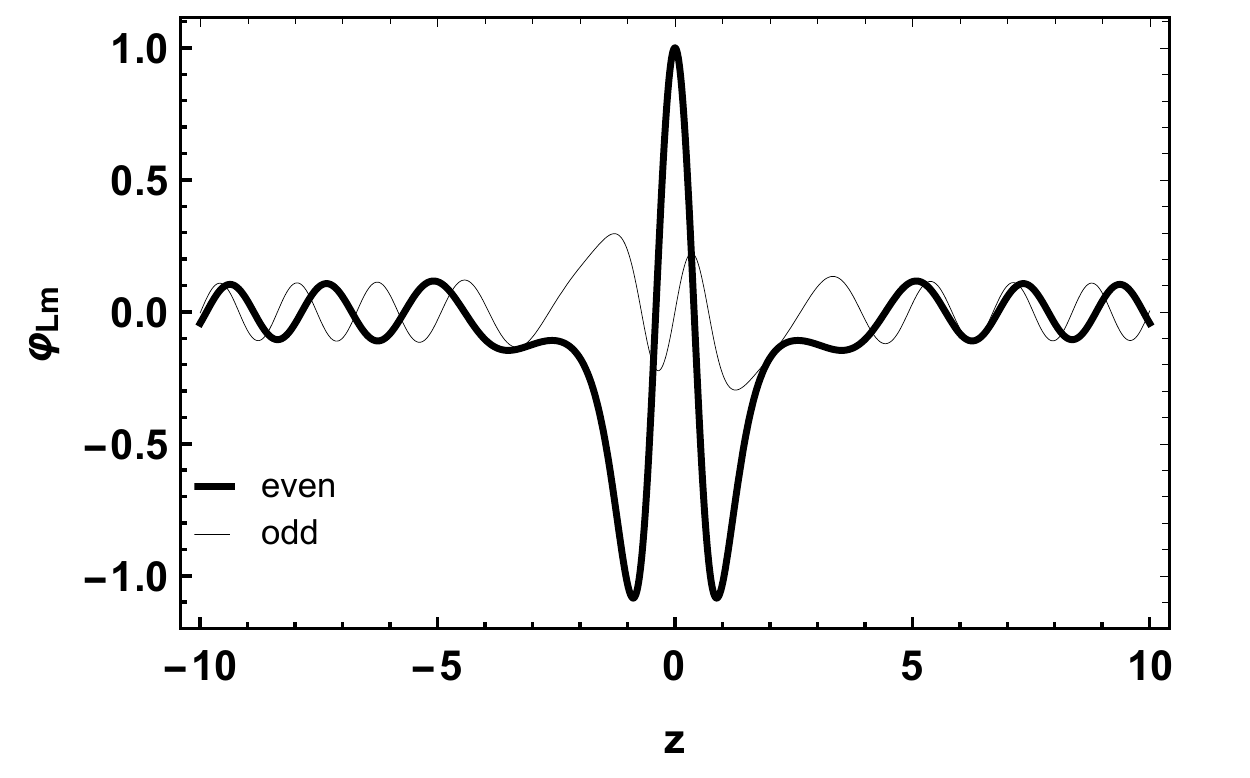}\\
(a)\hspace{6 cm}(b)\\
\includegraphics[height=5cm]{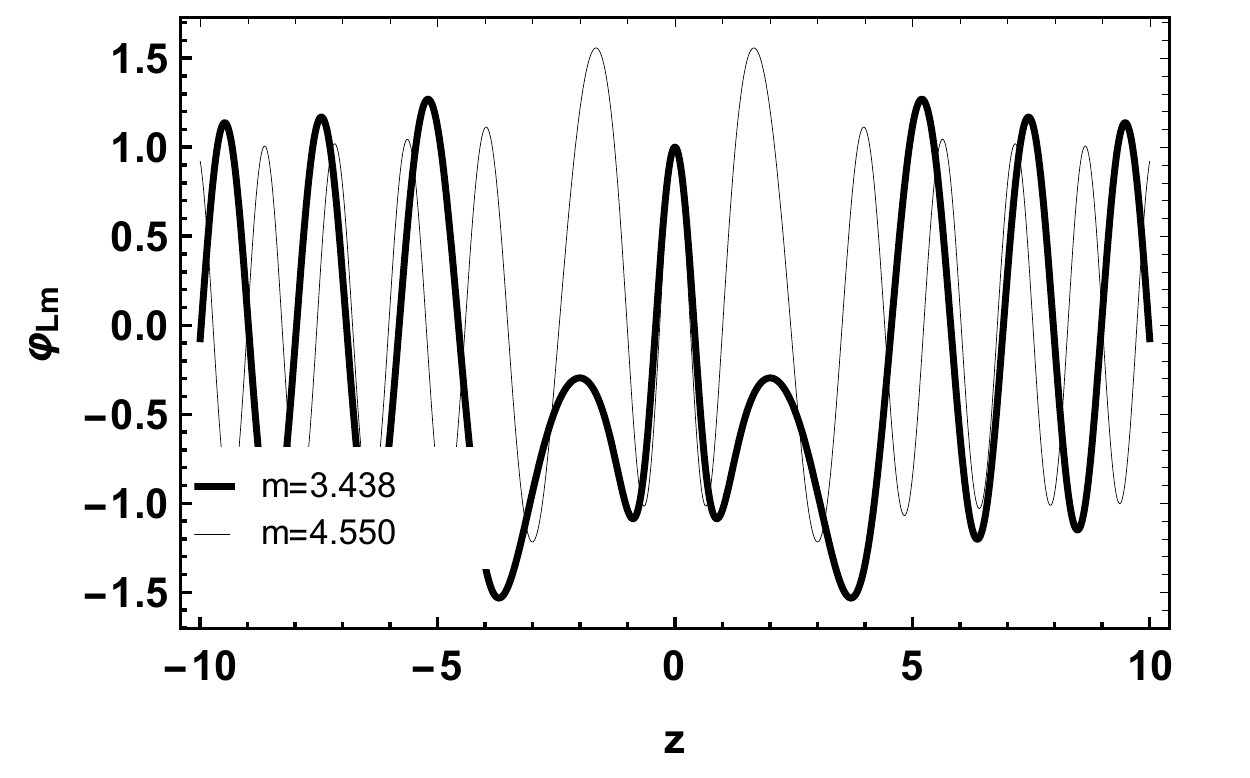} 
\includegraphics[height=5cm]{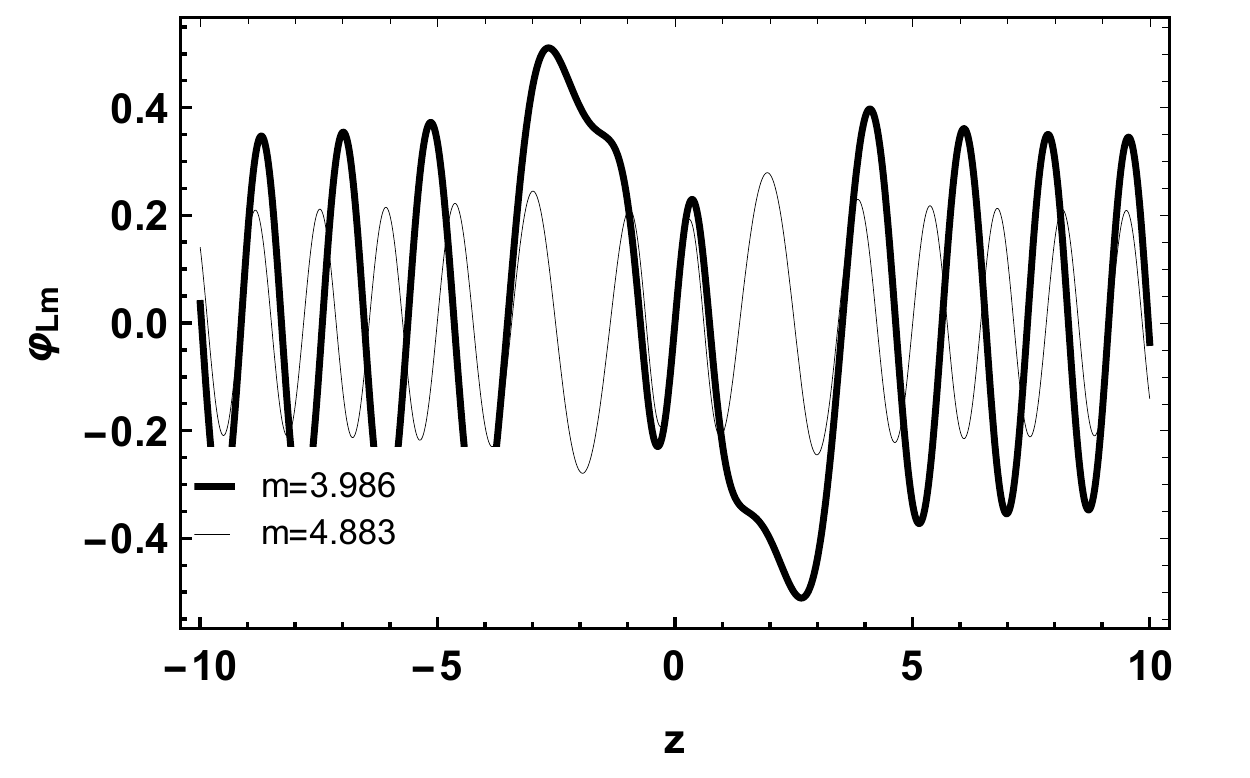}\\
(c) \hspace{6 cm}(d)\\
\includegraphics[height=5cm]{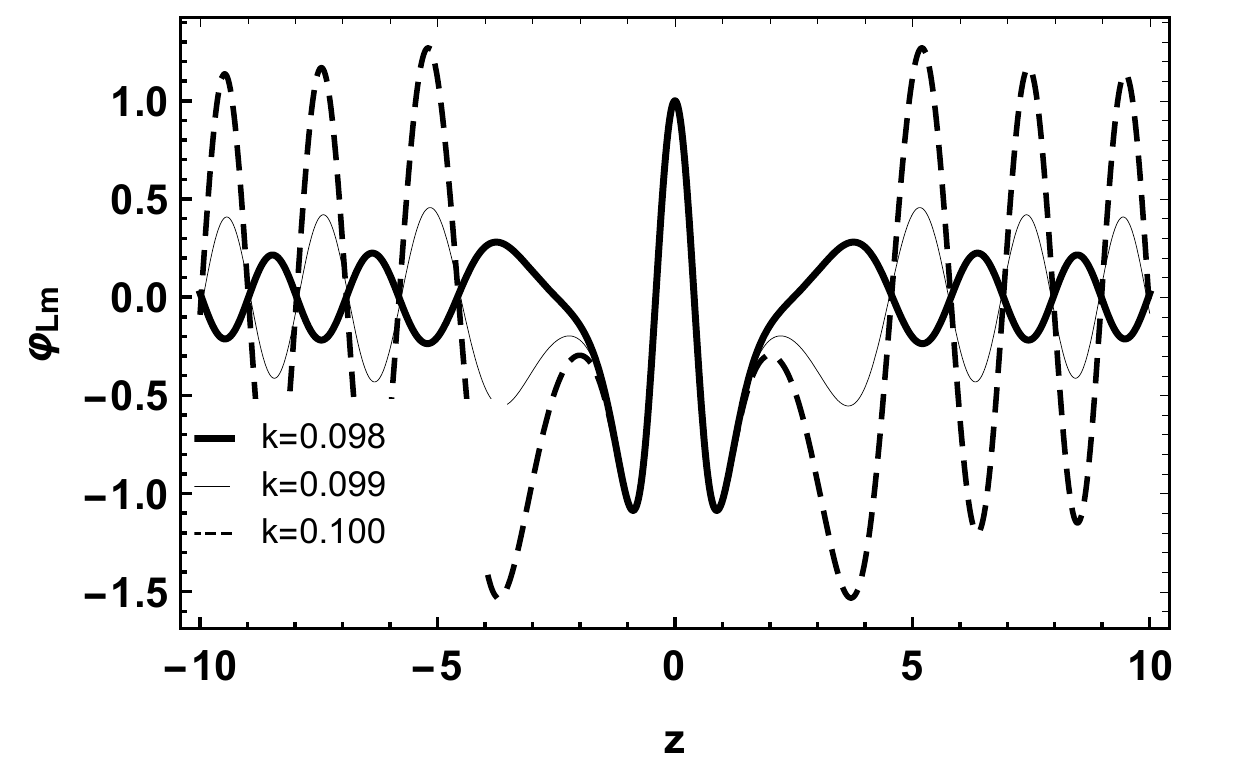} 
\includegraphics[height=5cm]{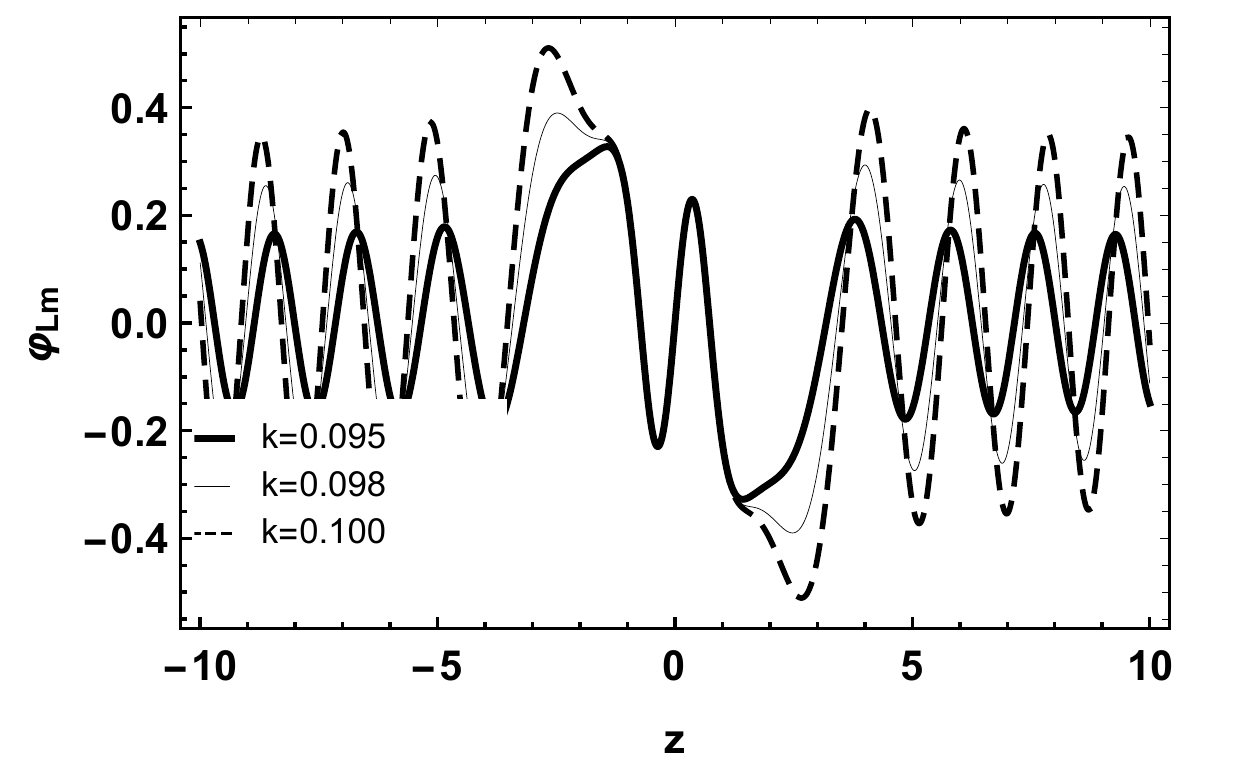}\\
(e) \hspace{6 cm}(f)
\end{tabular}
\end{center}
\caption{For $g_1(T)$, with $n_1=3$, $k=0.1$ and $p=\lambda=\xi=1$. (a) The form of the relative probability $P_L(m)$. (b) The resonant fermionic modes. The massive fermionic modes for $\varphi_{even}$ (c) and $\varphi_{odd}$ (d). By varying $k$, $\varphi_{even}$ with $m=3.438$ (e) and $\varphi_{odd}$ with $m=3.986$ (f).}
\label{fig7.2}
\end{figure}

In figure \ref{fig7.1}($a$) we show the form of the relative probability $P(m)$ for $g_1(T)$, with $n_1=2$. The peaks reveal the massive resonant states (Fig.\ref{fig7.1}$b$). In massive fermionic modes, for both even and odd solutions, we observe that when increasing the mass eigenvalue, the more oscillations we have, as can be seen in Figs.\ref{fig7.1}($c$) and \ref{fig7.1}($d$). On the other hand, from Figs.\ref{fig7.1}($e$) and  \ref{fig7.1}($f$), we observe that when decreasing the value of $k$, the smaller the amplitudes of the oscillations will be, but we have more oscillations. This becomes much more evident near the core of the brane.
We have something similar for the case of $n_1=3$, where the peaks of the relative probability $P(m)$ (Fig.\ref{fig7.2}$a$), reveal the massive resonant states (Fig.\ref{fig7.2}$b$). The greater the mass eigenvalue, the more oscillations we obtain (Figs.\ref{fig7.2}$c$ and \ref{fig7.2}$d$). Decreasing the value of $k$, smaller will be the amplitudes of the oscillations, mainly close to the brane core, as we can see in Figs.\ref{fig7.2}($e$) and \ref{fig7.2}($f$).

\begin{figure}[ht!]
\begin{center}
\begin{tabular}{ccc}
\includegraphics[height=5cm]{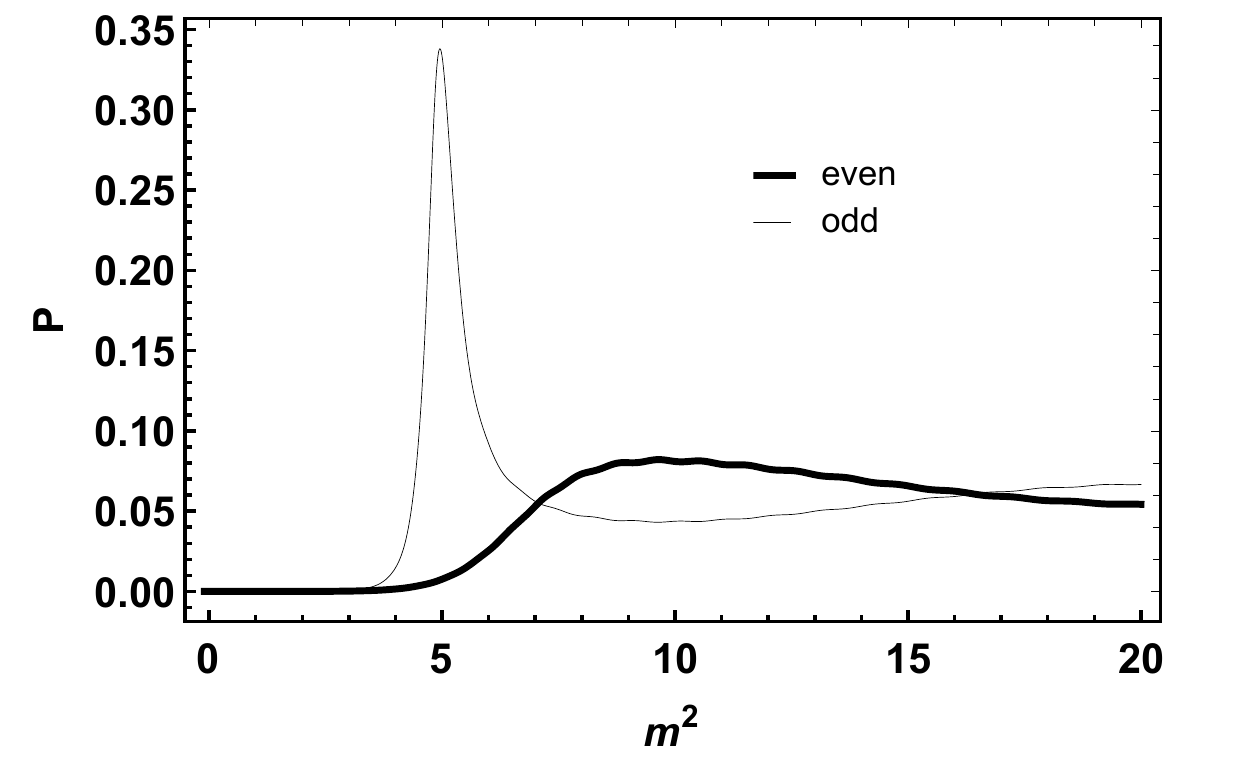}
\includegraphics[height=5cm]{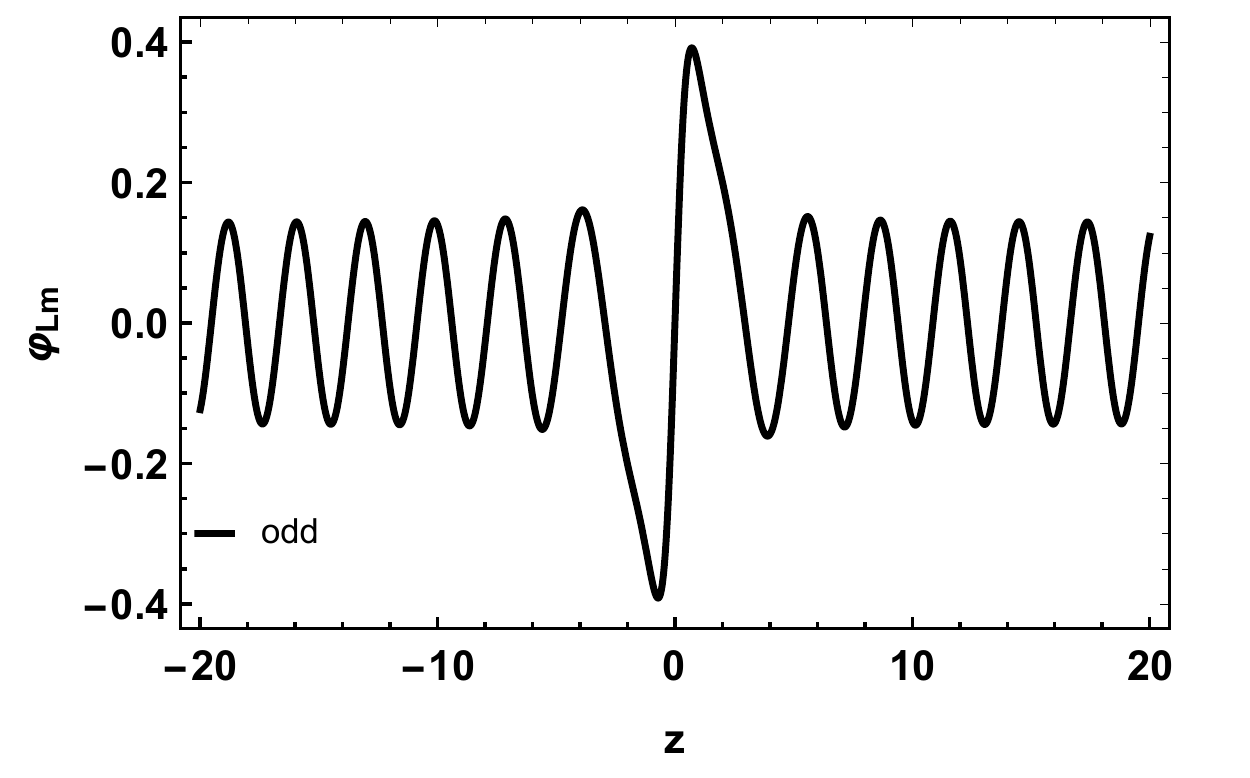}\\
(a)\hspace{6 cm}(b)\\
\includegraphics[height=5cm]{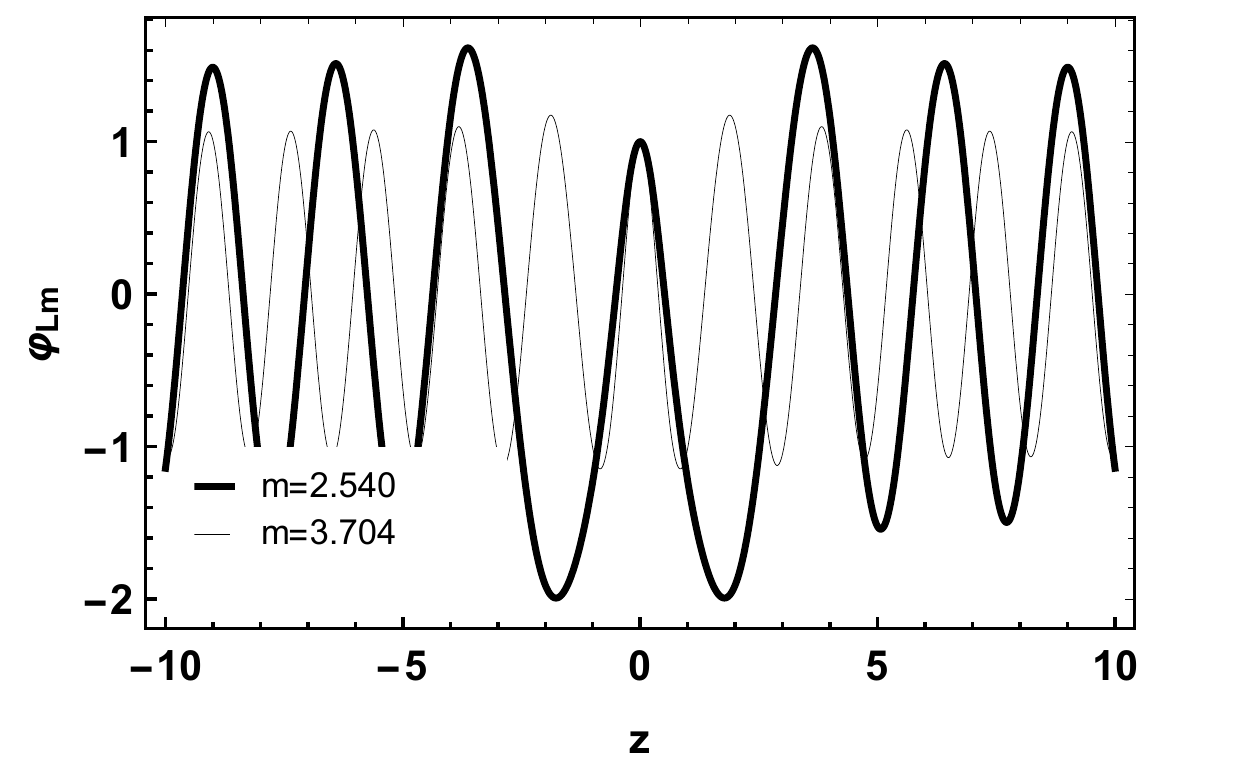} 
\includegraphics[height=5cm]{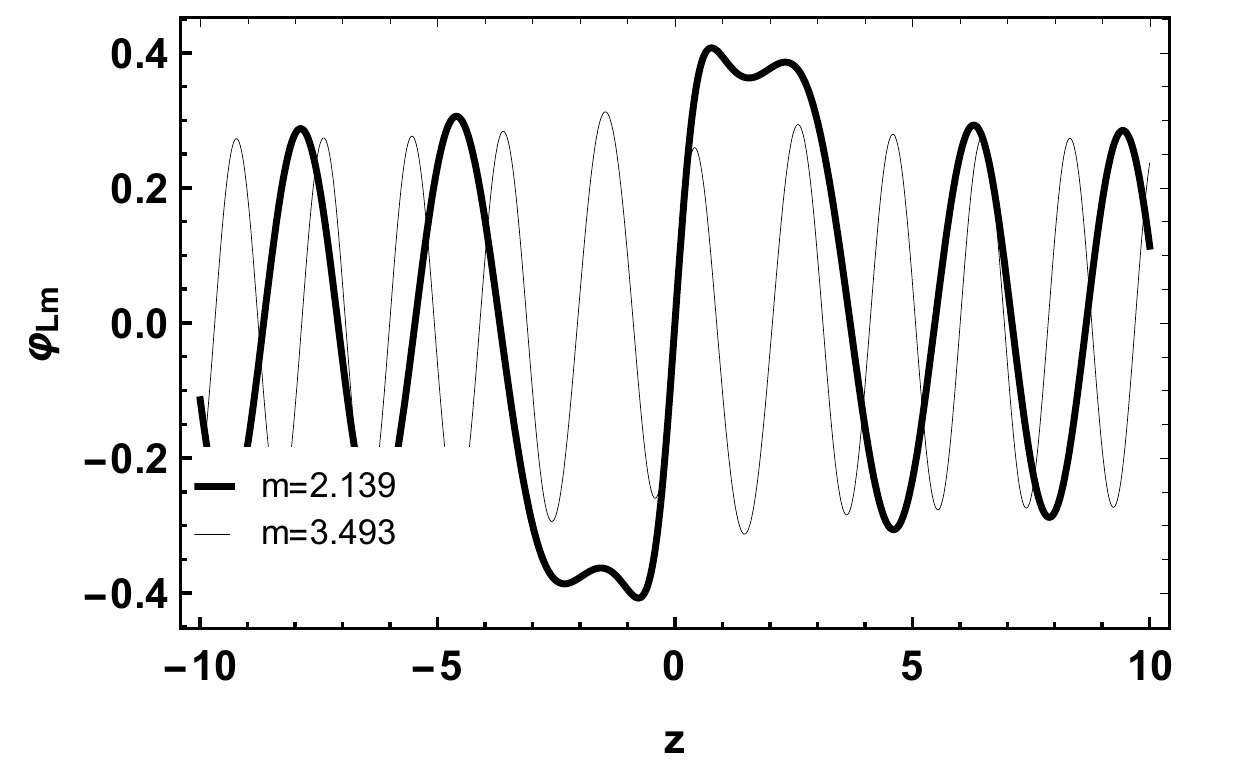}\\
(c) \hspace{6 cm}(d)\\
\includegraphics[height=5cm]{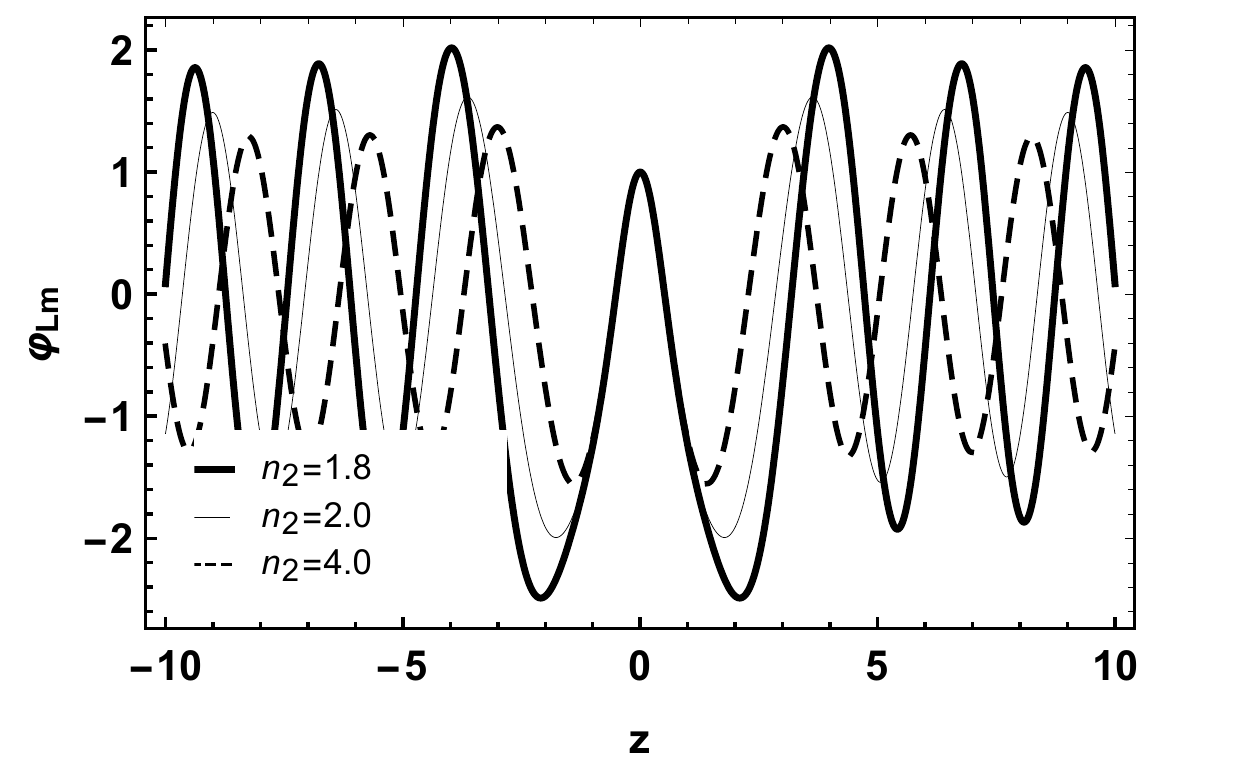} 
\includegraphics[height=5cm]{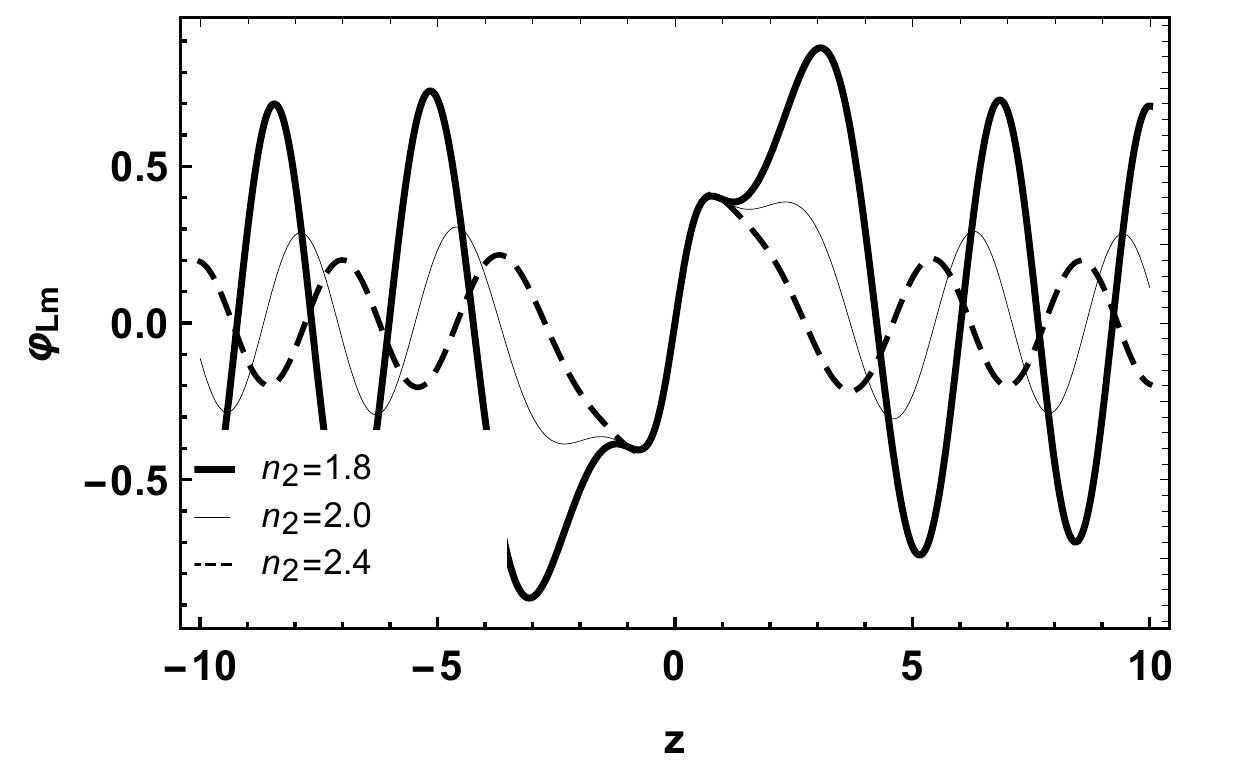}\\
(e) \hspace{6 cm}(f)
\end{tabular}
\end{center}
\caption{For $g_2(T)$, with $n_2=2$ and $p=\lambda=\xi=1$. (a) The form of the relative probability $P_L(m)$. (b) The resonant fermionic modes. The massive fermionic modes for $\varphi_{even}$ (c) and $\varphi_{odd}$ (d). By varying $n_2$, $\varphi_{even}$ with $m=2.540$ (e) and $\varphi_{odd}$ with $m=2.139$ (f).}
\label{fig8}
\end{figure}

In figure \ref{fig8}($a$) we show the form of the relative probability $P(m)$ for $g_2(T)$. We observe massive resonant states only in the odd solutions (Fig.\ref{fig8}$b$). In massive fermionic modes, for both even and odd solutions, we observe that when increasing the mass eigenvalue, the more oscillations we have, as can be seen in Figs.\ref{fig8}($c$) and \ref{fig8}($d$). When we increase the value of the torsional parameter $n_2$, we have more oscillations with smaller amplitudes, moving closer to the core of the brane, as can be seen in Figs.\ref{fig8}($e$) and  \ref{fig8}($f$).

\begin{figure}[ht!]
\begin{center}
\begin{tabular}{ccc}
\includegraphics[height=5cm]{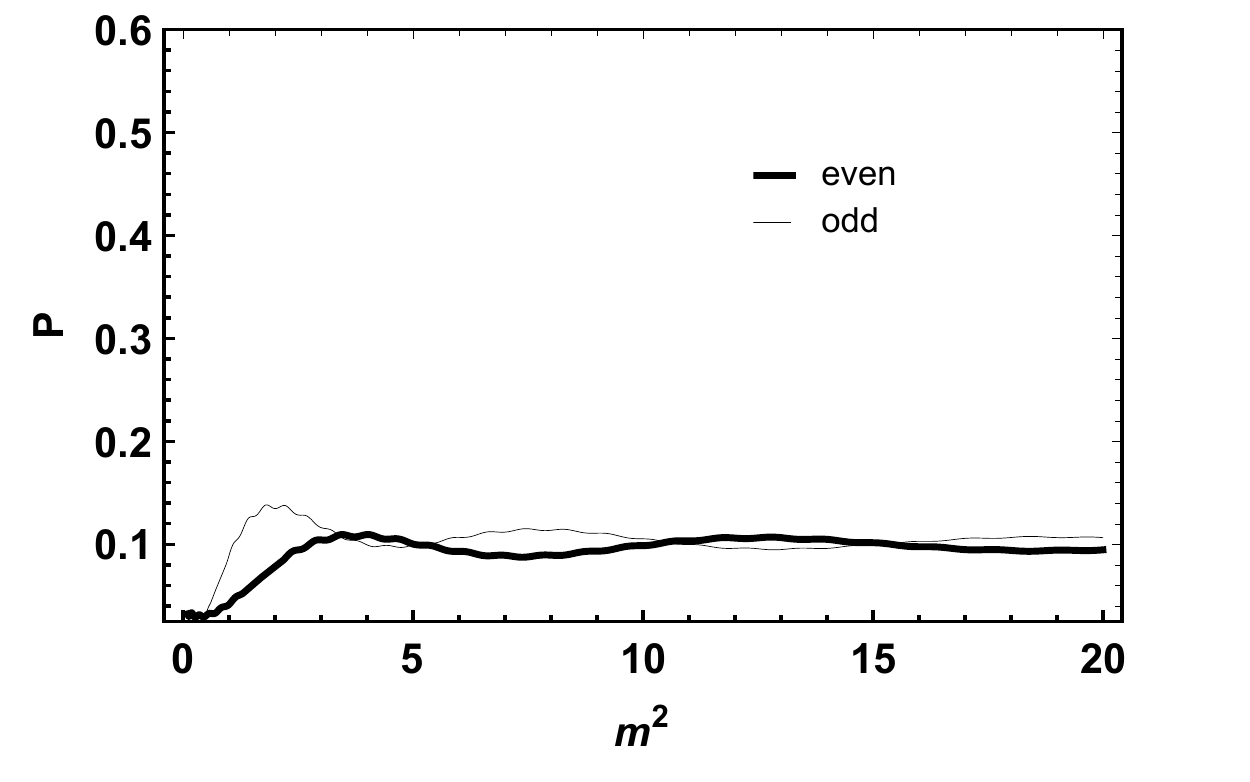}\\
(a)\\
\includegraphics[height=5cm]{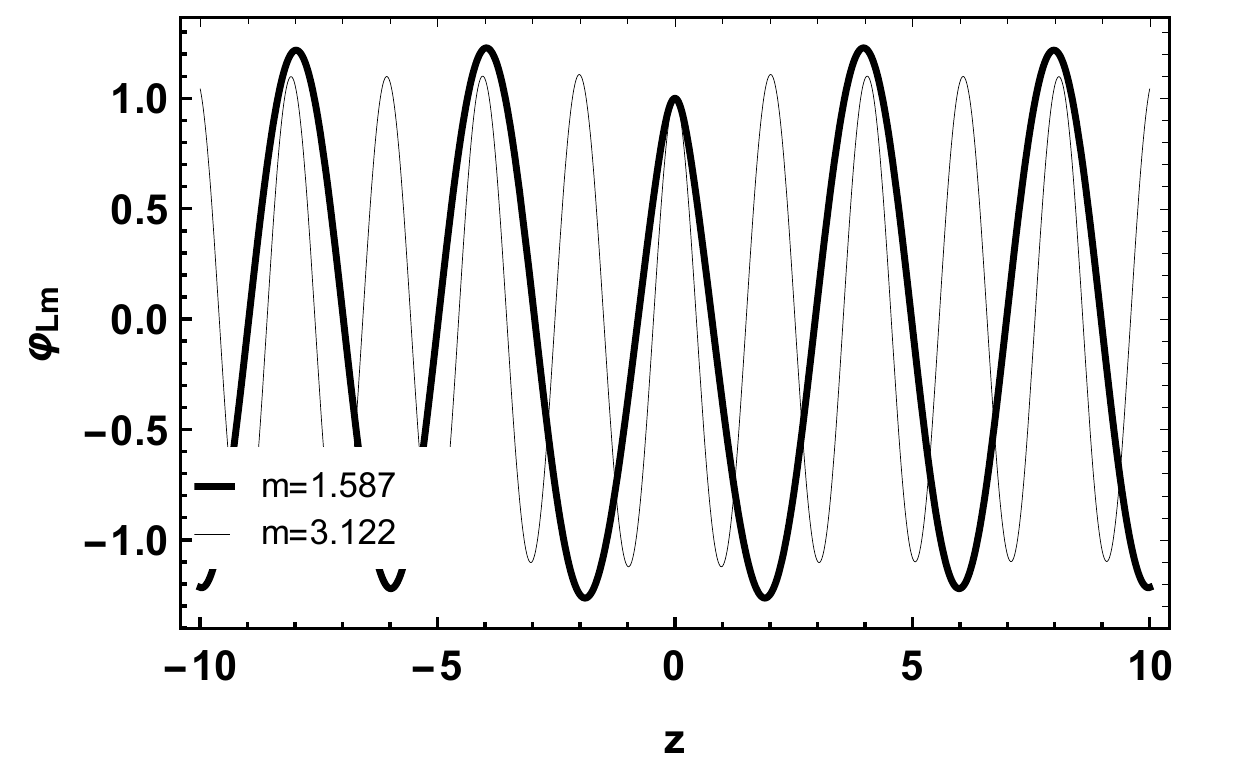} 
\includegraphics[height=5cm]{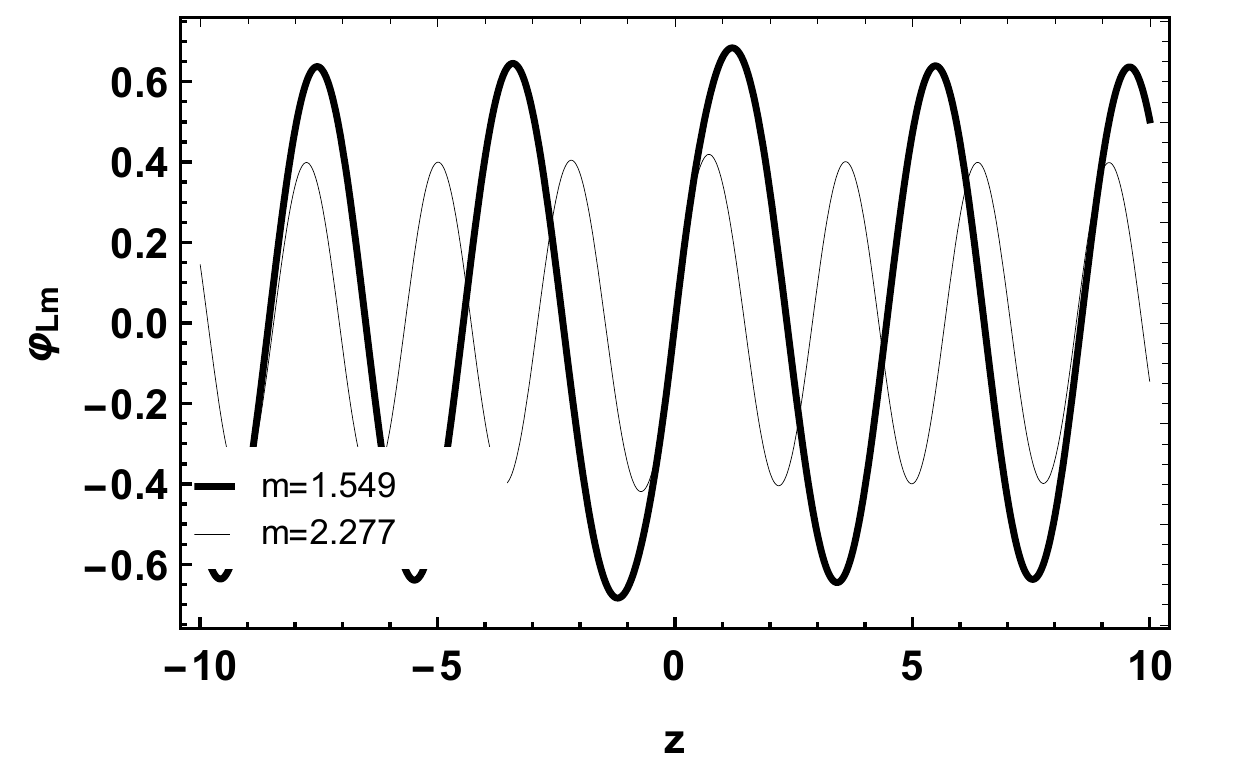}\\
(b) \hspace{6 cm}(c)\\
\includegraphics[height=5cm]{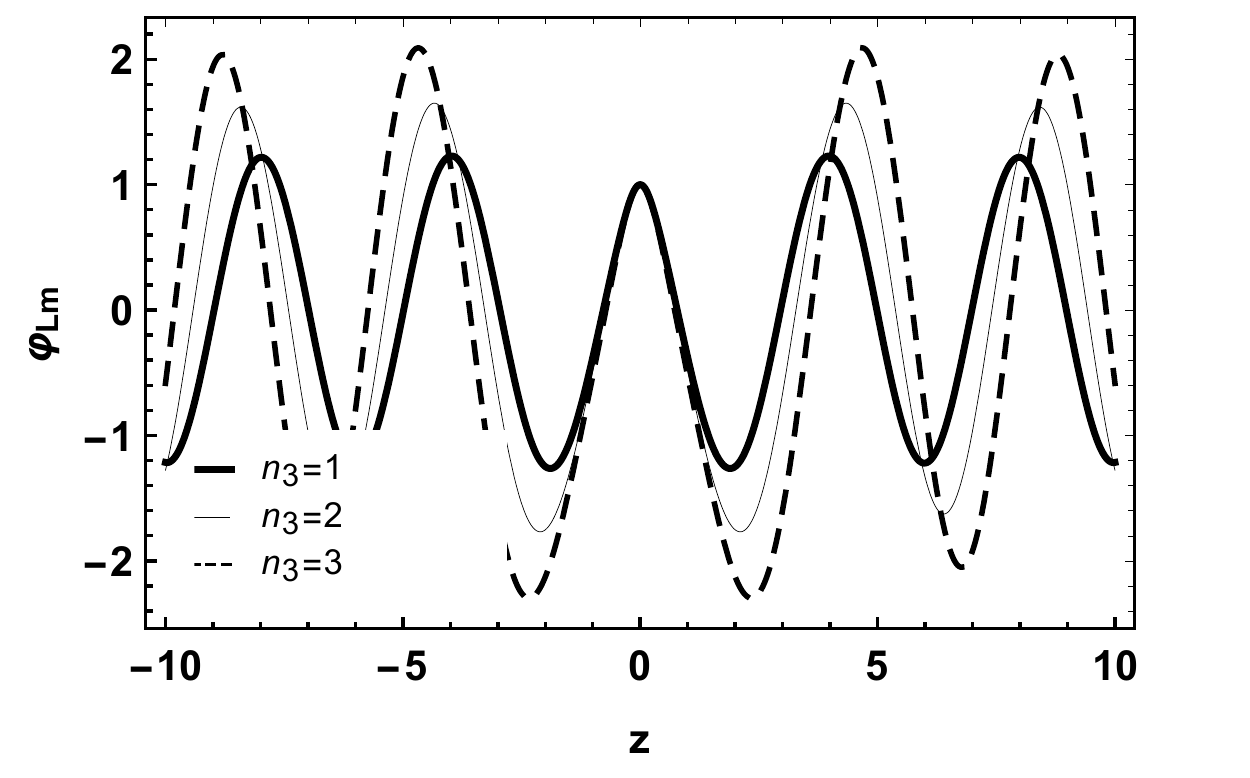} 
\includegraphics[height=5cm]{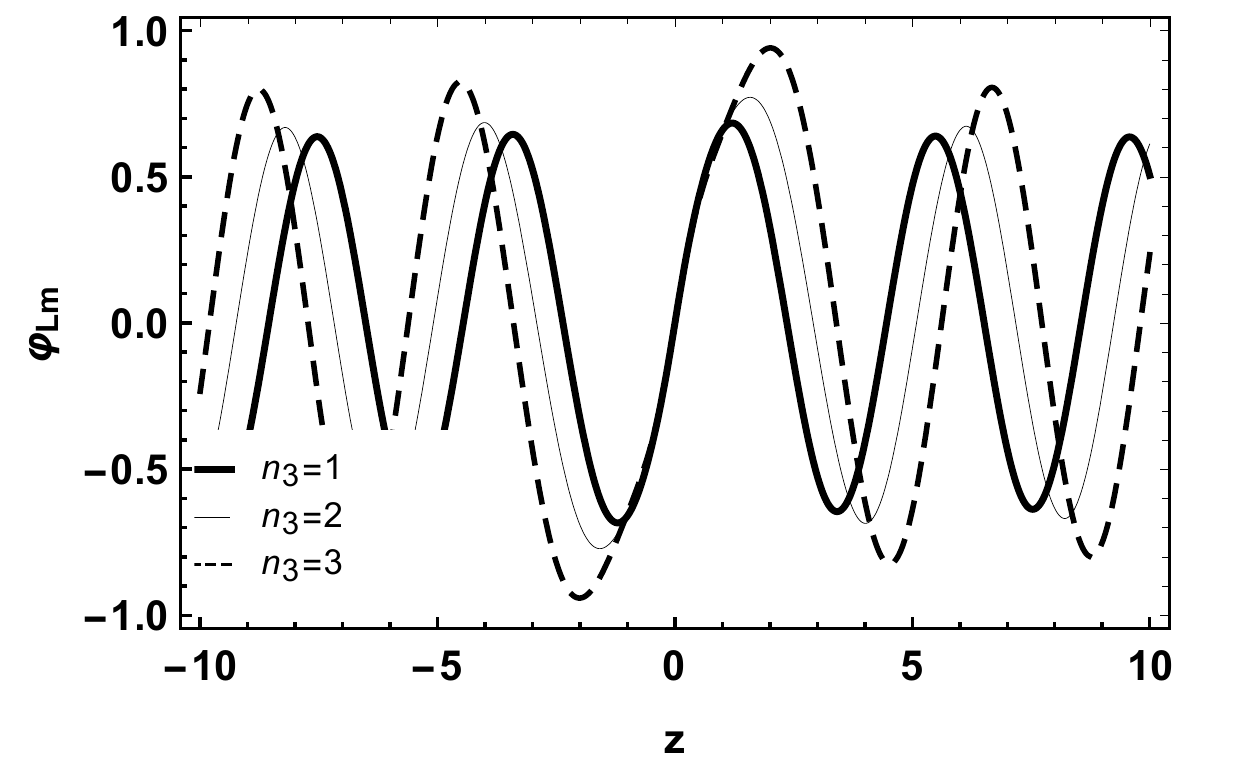}\\
(d) \hspace{6 cm}(e)
\end{tabular}
\end{center}
\caption{For $g_3(T)$, with $n_3=1$ and $p=\lambda=\xi=1$. (a) The form of the relative probability $P_L(m)$. The massive fermionic modes for $\varphi_{even}$ (b) and $\varphi_{odd}$ (c). By varying $n_3$, $\varphi_{even}$ with $m=1.587$ (d) and $\varphi_{odd}$ with $m=1.549$ (e).}
\label{fig9}
\end{figure}

For $g_3(T)$ the relative probability $P(m)$ has no peaks, which indicates the absence of massive resonant states (Fig.\ref{fig9}$a$). In massive fermionic modes, for both even and odd solutions, we observe that when increasing the mass eigenvalue, the more oscillations we have. However, this occurs more subtly than in the previous cases, as can be seen in Figs.\ref{fig9}($b$) and \ref{fig9}($c$).
We obtain a behavior contrary to that obtained in the case $g_2(T)$, when we increase the value of the torsion parameter $n_3$, the greater are the amplitudes of the oscillations that move away from the brane core (Figs.\ref{fig9}$d$ and \ref{fig9}$e$).

\section{Final remarks}
\label{finalremarks}

There are many models of $f(T)$ gravity in the literature. These models are  used to explain outstanding physical problems, such as the acceleration of the universe, dark energy, and cosmology \cite{Linder,ftgw,ftenergyconditions}.
In this work we study three particular cases of these $f(T)$ models, in a braneworld scenario, namely, $f_1(T)=T+kT^{n_1}$, $f_2(T)=n_2\sinh\left(\frac{T}{ n_2}\right)$ and $f_3(T)=n_3\tanh\left(\frac{T}{n_3}\right)$, where $k$ and $n_{1,2,3}$ are the parameters that control the influence of the torsion.  We obtain the solution of the scalar field $\phi(y)$ for all cases $f_{1,2,3}(T)$. They assume a constant value $\phi_{c}$ asymptotically, the same happens to the potential that takes the form of a cosmological constant ($\Lambda\equiv V(\phi\rightarrow\pm\phi_{c})$). Therefore, we guarantee that the chosen models make physical sense. We observe that the parameters that control the torsion generate internal structures in the brane which tend to split it. This is evident with the double-kink type field solutions and from the energy densities. 

In order to localize fermions in branes one needs a coupling
between the spinors and the scalar fields that form the brane. In this work, we use a non-minimal coupling with the torsion and we obtained very good results. This achievement is particularly very interesting. It shows us that it is possible to use a purely geometric alternative to locate fermions in brane instead of the minimal Yukawa coupling. Furthermore, this non-minimal coupling can give us a simpler and more accurate analysis of the influence of torsion on the location of fermions in the brane. We considered a form of the non-minimal coupling $g(T)$ which is directly related to $f(T)$, namely, $g_1(T)=\sqrt{-f_1(T)}$ and $g_{2,3}(T)=f_{2,3}(\sqrt{-T})$.

For both cases $g_{1}(T)$ and $g_{2,3}(T)$, only left-chiral fermions are located. We note that the parameters that control torsion affect directly the behavior of the effective potentials $V_{L,R}$, also modifying the massless fermionic modes. As expected, massive fermionic modes are also affected by the parameters that control torsion, which show an asymptotic divergence, forming non-localized states. This behavior is characteristic of free modes, typical of plane wave oscillations. This clearly shows that these massive fermions will be leaked from the brane. Similar results were obtained in Refs. \cite{Liu:2008pi,Moreira20211}. This behavior was already expected, since the effective potential vanishes asymptotically.

In addition,  an interesting result was the sharper chirality localization of the bulk fermion on the brane depending on the parameters $k$ and $n_{1,2,3}$ of the functions $g_{1,2,3}(T)$. In the Standard Model of particles, the electron has left-handed chirality. Usually, warped braneworld models lead to the localization of the left-handed spin $1/2$ fermion on the brane while the right-handed fermion is not confined on the brane. Thus, the localization of only the left-handed fermion can be used to rule out those configurations where only the right-handed spinor is confined at the brane. Therefore, non-minimal coupling to torsion has an observational viability equivalence to the theory of the Standard Model of particles.

\section*{Acknowledgments}
\hspace{0.5cm}The authors thank the Conselho Nacional de Desenvolvimento Cient\'{\i}fico e Tecnol\'{o}gico (CNPq), grants n$\textsuperscript{\underline{\scriptsize o}}$ 312356/2017-0 (JEGS) and n$\textsuperscript{\underline{\scriptsize o}}$ 309553/2021-0 (CASA), and Coordenaçao de Aperfeiçoamento do Pessoal de Nível Superior (CAPES), for financial support. The authors also thank the anonymous referees for their valuable comments and suggestions.


\begin{thebibliography}{99}

\bibitem{Kaluza1921}
T.~Kaluza,
Sitzungsber. Preuss. Akad. Wiss. Berlin (Math. Phys. ) \textbf{1921}, 966-972 (1921).

\bibitem{Klein1926}
O.~Klein,
Nature \textbf{118}, 516 (1926).

\bibitem{Arkani-Hamed:1998jmv}
N.~Arkani-Hamed, S.~Dimopoulos and G.~R.~Dvali,
Phys. Lett. B \textbf{429}, 263-272 (1998).

\bibitem{rs}
    L.~Randall and R.~Sundrum, Phys.\ Rev.\ Lett.\  {\bf 83}, 4690 (1999). 
        
\bibitem{rs2}
    L.~Randall and R.~Sundrum, Phys.\ Rev.\ Lett.\  {\bf 83}, 3370 (1999).

\bibitem{cosmologicalconstant}
J.~M.~Schwindt and C.~Wetterich,
  Nucl.\ Phys.\ B {\bf 726}, 75 (2005).

\bibitem{darkmatter}
  T.~Gherghetta and B.~von Harling,
  JHEP {\bf 1004}, 039 (2010).
 
\bibitem{Giddings2001}
S.~B.~Giddings and S.~D.~Thomas,
Phys. Rev. D \textbf{65}, 056010 (2002).

\bibitem{Gregory:2000jc}
R.~Gregory, V.~A.~Rubakov and S.~M.~Sibiryakov,
Phys. Rev. Lett. \textbf{84}, 5928-5931 (2000).

\bibitem{Appelquist:2000nn}
T.~Appelquist, H.~C.~Cheng and B.~A.~Dobrescu,
Phys. Rev. D \textbf{64}, 035002 (2001).

\bibitem{Dvali:2000hr}
G.~R.~Dvali, G.~Gabadadze and M.~Porrati,
Phys. Lett. B \textbf{485}, 208-214 (2000).

\bibitem{DeWolfe:1999cp}
O.~DeWolfe, D.~Z.~Freedman, S.~S.~Gubser and A.~Karch,
Phys. Rev. D \textbf{62}, 046008 (2000).

\bibitem{Gremm1999}
M.~Gremm,
Phys. Lett. B \textbf{478}, 434 (2000).

\bibitem{Csakil}
C.~Csaki, J.~Erlich, T.~J.~Hollowood and Y.~Shirman,
Nucl. Phys. B \textbf{581}, 309-338 (2000).

\bibitem{Gremm2000}
M.~Gremm,
Phys. Rev. D \textbf{62}, 044017 (2000).

\bibitem{Dzhunushaliev:2009}
V.~Dzhunushaliev, V.~Folomeev and M.~Minamitsuji,
Rept. Prog. Phys. \textbf{73}, 066901 (2010).

\bibitem{Dzhunushaliev:2010}
V.~Dzhunushaliev and V.~Folomeev,
Gen. Rel. Grav. \textbf{43}, 1253-1261 (2011).

\bibitem{Herrera-Aguilar:2009}
A.~Herrera-Aguilar, D.~Malagon-Morejon, R.~R.~Mora-Luna and U.~Nucamendi,
Mod. Phys. Lett. A \textbf{25}, 2089-2097 (2010).

\bibitem{Dzhunushaliev:2007}
V.~Dzhunushaliev, V.~Folomeev, D.~Singleton and S.~Aguilar-Rudametkin,
Phys. Rev. D \textbf{77}, 044006 (2008).

\bibitem{Gogberashvili:2003a}
M.~Gogberashvili and D.~Singleton,
Phys. Rev. D \textbf{69}, 026004 (2004).

\bibitem{Gogberashvili:2003b}
M.~Gogberashvili and D.~Singleton,
Phys. Lett. B \textbf{582}, 95-101 (2004).

\bibitem{Rubakov:1983bb}
V.~A.~Rubakov and M.~E.~Shaposhnikov,
Phys. Lett. B \textbf{125}, 136-138 (1983).

\bibitem{Goldberger1999}
W.~D.~Goldberger and M.~B.~Wise,
Phys. Rev. Lett. \textbf{83} (1999), 4922.


\bibitem{Bazeia2008}
D.~Bazeia, A.~R.~Gomes, L.~Losano and R.~Menezes,
Phys. Lett. B \textbf{671} (2009), 402.

\bibitem{Geng:2015kvs}
W.~J.~Geng and H.~Lu,
Phys. Rev. D \textbf{93}, no.4, 044035 (2016).

\bibitem{Dzhunushaliev:2011mm}
V.~Dzhunushaliev and V.~Folomeev,
Gen. Rel. Grav. \textbf{44}, 253-261 (2012).


\bibitem{Kehagias}
  A.~Kehagias and K.~Tamvakis,
Phys.\ Lett.\ B {\bf 504}, 38 (2001).


\bibitem{Almeida2009}
C.~A.~S.~Almeida, M.~M.~Ferreira, Jr., A.~R.~Gomes and R.~Casana,
Phys. Rev. D \textbf{79}, 125022 (2009).


\bibitem{RandjbarDaemi2000}
S.~Randjbar-Daemi and M.~E.~Shaposhnikov,
Phys. Lett. B \textbf{492}, 361-364 (2000).

\bibitem{Liu2009}
Y.~X.~Liu, J.~Yang, Z.~H.~Zhao, C.~E.~Fu and Y.~S.~Duan,
Phys. Rev. D \textbf{80}, 065019 (2009).

\bibitem{Liu2009a}
Y.~X.~Liu, H.~T.~Li, Z.~H.~Zhao, J.~X.~Li and J.~R.~Ren,
JHEP \textbf{10}, 091 (2009).

\bibitem{Liu2008}
Y.~X.~Liu, L.~D.~Zhang, L.~J.~Zhang and Y.~S.~Duan,
Phys. Rev. D \textbf{78}, 065025 (2008).



\bibitem{Liu2009b}
Y.~X.~Liu, C.~E.~Fu, L.~Zhao and Y.~S.~Duan,
Phys. Rev. D \textbf{80}, 065020 (2009).

\bibitem{Liu2008b}
Y.~X.~Liu, L.~D.~Zhang, S.~W.~Wei and Y.~S.~Duan,
JHEP \textbf{08}, 041 (2008).

\bibitem{Obukhov2002}
Y.~N.~Obukhov and J.~G.~Pereira,
Phys. Rev. D \textbf {67}, 044016 (2003)



\bibitem{Ulhoa2016}
S.~C.~Ulhoa, A.~F.~Santos and F.~C.~Khanna,
Gen. Rel. Grav. {\bf 49}, 54 (2017).

\bibitem{Dantas2013}
D.~M.~Dantas, J.~E.~G.~Silva and C.~A.~S.~Almeida,
Phys. Lett. B \textbf {725}, 425-430 (2013).
  
\bibitem{Sousa2014}
L.~J.~S.~Sousa, C.~A.~S.~Silva, D.~M.~Dantas and C.~A.~S.~Almeida,
Phys. Lett. B \textbf {731}, 64-69 (2014).
  
\bibitem{Dantas}
D.~M.~Dantas, D.~F.~S.~Veras, J.~E.~G.~Silva and C.~A.~S.~Almeida,
Phys. Rev. D {\bf 92}, 104007 (2015).


\bibitem{Mitra2017}
J.~Mitra, T.~Paul and S.~SenGupta,
EUR. Phys. J. C \textbf {77}, 833 (2017).

\bibitem{Buyukdag2018}
Y.~Buyukdag, T.~Gherghetta and A.~S.~Miller,
Phys. Rev. D \textbf {99}, no.3, 035046 (2019).

\bibitem{Wang2019}
L.~L.~Wang, H.~Guo, C.~E.~Fu and Q.~Y.~Xie,
 `` Gravity and Matters on a pure geometric thick polynomial f(R) brane, ''


\bibitem{ft}
Y.~F.~Cai, S.~Capozziello, M.~De Laurentis and E.~N.~Saridakis,
Rept. Prog. Phys. \textbf{79}, no.10, 106901 (2016).

\bibitem{Ferraroinflation}
R.~Ferraro and F.~Fiorini,
Phys. Rev. D \textbf{75}, 084031 (2007).

\bibitem{Andrade2001}
V.~C.~de Andrade, L.~C.~T.~Guillen and J.~G.~Pereira,
Phys. Rev. D \textbf {64}, 027502 (2001).

\bibitem{Aldrovandi}
R. Aldrovandi and J. G. Pereira, \textit{Teleparallel Gravity: An Introduction},
(Springer, Berlin, 2013).

\bibitem{Linder}
E.~V.~Linder,
Phys. Rev. D \textbf{81}, 127301 (2010).


\bibitem{Bahamonde:2021gfp}
S.~Bahamonde, K.~F.~Dialektopoulos, C.~Escamilla-Rivera, G.~Farrugia, V.~Gakis, M.~Hendry, M.~Hohmann, J.~L.~Said, J.~Mifsud and E.~Di Valentino,
``Teleparallel Gravity: From Theory to Cosmology,''


\bibitem{ftpalatini}
J.~Beltr\'{a}n Jim\'{e}nez, L.~Heisenberg and T.~S.~Koivisto,
JCAP \textbf{08}, 039 (2018).


\bibitem{Yang2017}
K.~Yang, W.~D.~Guo, Z.~C.~Lin and Y.~X.~Liu,
Phys. Lett. B \ textbf {782}, 170-175 (2018).

\bibitem{ftenergyconditions}
D.~Liu and M.~Reboucas,
Phys. Rev. D \textbf{86}, 083515 (2012).
  
\bibitem{tensorperturbations}
W.~D.~Guo, Q.~M.~Fu, Y.~P.~Zhang and Y.~X.~Liu,
Phys. Rev. D \textbf{93}, no.4, 044002 (2016).

\bibitem{Yang2012}
J.~Yang, Y.~-L.~Li, Y.~Zhong and Y.~Li,
  Phys.\ Rev.\ D {\bf 85}, 084033 (2012).    

\bibitem{Ferraro2011us}
R.~Ferraro and F.~Fiorini,
Phys. Lett. B \textbf{702}, 75-80 (2011).

\bibitem{Tamanini2012}
N.~Tamanini and C.~G.~Boehmer,
Phys. Rev. D \textbf{86}, 044009 (2012).



\bibitem{Menezes}
R.~Menezes,
Phys. Rev. D \textbf{89}, no.12, 125007 (2014).

\bibitem{ftnoncanonicalscalar}
J.~Wang, W.~D.~Guo, Z.~C.~Lin and Y.~X.~Liu,
Phys. Rev. D \textbf{98}, no.8, 084046 (2018).

\bibitem{ftborninfeld}
K.~Yang, W.~D.~Guo, Z.~C.~Lin and Y.~X.~Liu,
Phys. Lett. B \textbf{782}, 170-175 (2018).

\bibitem{ftmimetic}
W.~D.~Guo, Y.~Zhong, K.~Yang, T.~T.~Sui and Y.~X.~Liu,
Phys. Lett. B \textbf{800}, 135099 (2020).

\bibitem{ftgw}
K.~Bamba, S.~Capozziello, M.~De Laurentis, S.~Nojiri and D.~S\'{a}ez-G\'{o}mez,
Phys. Lett. B \textbf{727}, 194-198 (2013).

\bibitem{mirza} B. Mirza and F. Oboudiat, JCAP {\bf 11},  011 (2017).


  
\bibitem{Arias2002ew}
O.~Arias, R.~Cardenas and I.~Quiros,
Nucl. Phys. B \textbf{643}, 187-200 (2002).

\bibitem{Barbosa-Cendejas2005vog}
N.~Barbosa-Cendejas and A.~Herrera-Aguilar,
JHEP \textbf{10}, 101 (2005).

\bibitem{Barbosa-Cendejas:2006cic}
N.~Barbosa-Cendejas and A.~Herrera-Aguilar,
Phys. Rev. D \textbf{73}, 084022 (2006).

\bibitem{Liu:2008pi}
Y.~X.~Liu, L.~D.~Zhang, L.~J.~Zhang and Y.~S.~Duan,
Phys. Rev. D \textbf{78}, 065025 (2008).

\bibitem{Liu:2009uca}
Y.~X.~Liu, H.~Guo, C.~E.~Fu and J.~R.~Ren,
JHEP \textbf{02}, 080 (2010).

\bibitem{Barbosa-Cendejas:2007cwl}
N.~Barbosa-Cendejas, A.~Herrera-Aguilar, M.~A.~Reyes Santos and C.~Schubert,
Phys. Rev. D \textbf{77}, 126013 (2008).

\bibitem{Liu:2009dwa}
Y.~X.~Liu, Z.~H.~Zhao, S.~W.~Wei and Y.~S.~Duan,
JCAP \textbf{02}, 003 (2009).

\bibitem{Alencar:2012en}
G.~Alencar, R.~R.~Landim, M.~O.~Tahim and R.~N.~C.~Filho,
JHEP \textbf{01}, 050 (2013).

\bibitem{Moreira20211}
A.~R.~P.~Moreira, J.~E.~G.~Silva and C.~A.~S.~Almeida,
Eur. Phys. J. C \textbf{81}, no.4, 298 (2021).
  
\end{thebibliography}

\end{document}